\documentclass[aps,prb,twocolumn, showpacs, superscriptaddress]{revtex4-1}
\usepackage[dvips]{graphicx}
\usepackage[usenames,dvipsnames]{color}
\usepackage{amsmath}

\newcommand{\bra}[1] {\langle #1|}
\newcommand{\ket}[1] {| #1 \rangle}
\newcommand{\e} {\mathrm{e}}
\newcommand{\ii} {\mathrm{i}}

\begin{document}

\title{The Optical Excitation of Zigzag Carbon Nanotubes with Photons Guided in Nanofibers}

\author{S.~Broadfoot}
\affiliation{Clarendon Laboratory, University of Oxford, Parks Road, Oxford OX1 3PU, United Kingdom}
\author{U.~Dorner}
\affiliation{Centre for Quantum Technologies, National University of Singapore, 117543, Singapore}
\affiliation{Clarendon Laboratory, University of Oxford, Parks Road, Oxford OX1 3PU, United Kingdom}
\author{D.~Jaksch}
\affiliation{Clarendon Laboratory, University of Oxford, Parks Road, Oxford OX1 3PU, United Kingdom}
\affiliation{Centre for Quantum Technologies, National University of Singapore, 117543, Singapore}

\date{\today} 

\begin{abstract}
  We consider the excitation of electrons in semiconducting carbon
  nanotubes by photons from the evanescent field created by a
  subwavelength-diameter optical fiber. The strongly changing
  evanescent field of such nanofibers requires dropping the dipole
  approximation. We show that this leads to novel effects, especially
  a high dependence of the photon absorption on the relative
  orientation and geometry of the nanotube-nanofiber setup in the
  optical and near infrared domain. In particular, we calculate photon
  absorption probabilities for a straight nanotube and nanofiber
  depending on their relative angle. Nanotubes orthogonal to the fiber
  are found to perform much better than parallel nanotubes when they
  are short. As the nanotube gets longer the absorption of parallel
  nanotubes is found to exceed the orthogonal nanotubes and approach
  100\% for extremely long nanotubes. In addition, we show that if the
  nanotube is wrapped around the fiber in an appropriate way the
  absorption is enhanced. We find that optical and near infrared
  photons could be converted to excitations with efficiencies that may exceed
  90\%. This may provide opportunities for future photodetectors and we discuss possible setups.
\end{abstract}

\maketitle

\section{Introduction \label{sec1}}
The unique physical properties of carbon nanotubes and the flexibility
they provide in selecting their characteristics offers great potential
for nanotechnology~\cite{RevModPhys.79.677,JPSJ.74.777,saito1998physical}. Carbon
nanotubes can be either semi-conducting or metallic, depending on
their diameter and helical configuration. They typically have
nanometer sized diameters and a length of a few microns, although
centimetre long nanotubes have been
produced recently~\cite{doi:10.1021/nl901260b}. This makes them ideal 1D
systems that possess a ballistic conducting
channel~\cite{ISI:000073761000044}, no backward scattering, and energy
levels that can be adjusted with external
fields~\cite{PhysRevLett.106.156809,PhysRevB.64.153404,0953-8984-17-37-019}.
Superconductivity has also been observed in multi-walled carbon
nanotubes and single carbon nanotubes have exhibited a superconducting
proximity
effect~\cite{PhysRevLett.96.057001,PhysRevB.83.165420,Kasumov28051999,Morpurgo08101999}.
Their applications range from extremely strong fibers and organic
electronics~\cite{PhysRevB.79.085402} to electrochemical sensors~\cite{Snow25032005,doi:10.1021/nl060613v} and photon
detectors~\cite{doi:10.1021/nl034313e}.

\begin{figure}[t]
  \centering\includegraphics[width=8cm]{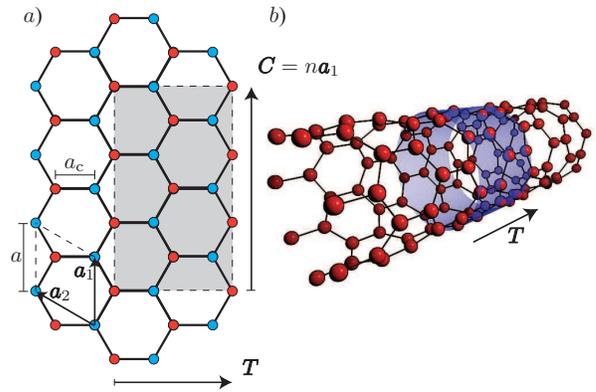}
  \caption{%
    (color online) a) Graphene lattice with the unit cells vectors labeled $\mathbf{a}_1$ and $\mathbf{a}_2$. These vectors have the length $a$.
    Here the atoms in the A sublattice are red (dark grey) and the B
    sublattice is cyan (light grey). $a_c$ is the distance between neighboring atoms. The unwrapped unit cell for a
    zigzag (3,0) nanotube is shown shaded and the \textbf{C} vector defines the
    nanotubes circumference. b) A section from a zigzag (7,0) nanotube
    is shown with its unit cell shaded. $\mathbf{T}$ is the tangential
    unit vector of the nanotube's unit cell.}
  \label{fig1}
\end{figure}

Carbon nanotubes are a form of carbon formed by rolling up a sheet of
graphene into a cylindrical tube. An illustration of this is given in
Fig.~\ref{fig1}. Here we focus on the optical properties of carbon
nanotubes. For a straight nanotube, inside a weak uniform classical
plane wave field, these properties have been extensively
studied~\cite{PhysRevLett.92.077402,Jiang20043169,PhysRevB.82.085442,PhysRevB.80.195422,PhysRevB.81.153402,PhysRevB.72.195403,PhysRevB.81.041414,PhysRevB.79.233406,PhysRevB.74.155434,PhysRevB.74.195431,ISI:000220192300002}.
Their quasi one-dimensionality means that their density of states
exhibit Van Hove singularities and these contribute to strong optical
absorption peaks. However, these results apply the dipole
approximation, where it is assumed that the field does not vary along
the nanotube's length. We extend this treatment by allowing for the
spatial dependence of the field. This situation is particularly
relevant when the electrons are delocalized in a tightly-confined
field, such that the field varies greatly over a few hundred
nanometres. The degree to which the electrons are delocalized is a
topic of ongoing research and various studies have been done on the
coherence length in nanotubes. Their results range from $10\,$nm to
several microns suggesting that the spatial field dependence is
certainly important for confined fields and may also be relevant for
plane
waves~\cite{PhysRevB.80.081410,PhysRevLett.95.076803,ISI:A1997WR25600045,PhysRevB.72.113410,PhysRevLett.94.086802}.
The systems we are primarily interested in are subwavelength-diameter
optical fibers coupled to carbon nanotubes. The electrical field of a
nanofiber is tightly confined and primarily exists outside of the
fiber, in a large evanescent field~\cite{Bures:99}. Due to
the presence of a strong field in a relativity small volume, these
nanofibers are ideal candidates to achieve a high optical absorption
in atomic systems. For example, their interaction with atom-arrays has
been studied~\cite{PhysRevA.72.063815,PhysRevLett.99.163602,PhysRevLett.104.203603}. However, in contrast to such
atom-fiber systems, the dipole approximation can not be applied in the
case of nanotubes since the optical field typically changes rapidly
along a nanotube's length. In this paper we calculate the (internal)
quantum efficiency, i.e. the probability that a nanotube, placed
inside the evanescent field of a nanofiber, absorbs a photon.
Calculations for the external quantum efficiency, i.e. the efficiency
for detecting the excitation with the photocurrent, are beyond the
scope of this paper. However, it should be noted that important
effects that could aid in this procedure, such as the avalanche
effect, have been observed in carbon
nanotubes~\cite{PhysRevLett.101.256804}. We focus specifically on the
example of zigzag nanotubes (see Fig.~\ref{fig1}) because they can be
direct semiconductors. However, our results are still representative
of other semiconducting nanotube types such as chiral nanotubes. We
find that the absorption is extremely dependent on the nanotube's
orientation. These results are highly relevant for the interface
between any future nanoscale photonics and carbon nanotubes. If the absorption process is coherent the system may also be suitable as a quantum memory, that maps a photonic quantum state on to a coherent excitation of the nanotube.

We will be using the band-to-band tight binding transition model for the carbon nanotube. This has proven itself to be very effective in determining the basic optical properties of nanotubes but does not include effects due to excitons~\cite{annurev.physchem.58.032806.104628} and electron-electron interactions~\cite{PhysRevB.80.153406}. Such effects give measurable corrections and there have been a few studies
considering the exciton absorption
strength~\cite{PhysRevLett.92.077402,doi:10.1021/nl0722525,PhysRevB.81.041414,JPSJ.68.3131}. Nevertheless, the band-to-band model is suitable to determine the main contributions to optical absorption.

This paper is organized as follows. We begin by giving an overview of nanotube properties and the calculation of their band structure in Sec.~\ref{sec2}. Based on this we then evaluate the photon absorption
by zigzag carbon nanotubes in Sec.~\ref{sec3}. In the setups
considered in this paper the nanotubes experience fields that change
strongly along their length, i.e. to calculate photon absorption we
can not rely on the dipole approximation. We obtain general
expressions for the absorption probability which are then applied to
cylindrical vacuum cladded silica nanofibers in Sec.~\ref{sec4}, and
discuss different geometrical setups of nanofiber and nanotube. Possible photodetectors that use these setups are then presented in Sec.~\ref{temp-deviceapp}. Finally, in Sec.~\ref{sec5}, we summarize our results.

\section{The Tight-binding Model for the Carbon Nanotube \label{sec2}}

Here we will review the basic properties of carbon nanotubes for completeness and layout the notation that we use in later sections. A single walled carbon nanotube (SWCNT) can be thought of as a sheet
of graphene wrapped into a tube, so we will start by describing the
tight-binding model of
graphene~\cite{doi:10.1080/00018732.2010.487978}. Graphene is a
regular 2D hexagonal Bravais lattice of carbon atoms and its structure
is shown in Fig.~\ref{fig1}. We label the unit vectors of the graphene
lattice $\mathbf{a}_1$ and $\mathbf{a}_2$. The length of these
vectors is the lattice constant $a$ which is related to the distance
between neighboring carbon atoms, $a_c$, by $a=\sqrt{3}a_c \simeq
0.246\,$nm.  Within each unit cell there are two carbon atoms, that we
label to form the A and B sublattices. We can then define the unit
vectors of the reciprocal lattice as $\mathbf{b}_1$ and
$\mathbf{b}_2$, with $\mathbf{a}_i.\mathbf{b}_j=2\pi \delta_{i j}$. The first Brillouin zone given by these is also
hexagonal. It has a selection of points with high-symmetry; one at the
center, the midpoints of the hexagonal edges and two
inequivalent types of corners.

\begin{figure}[ht]
  \centering\includegraphics[width=8cm]{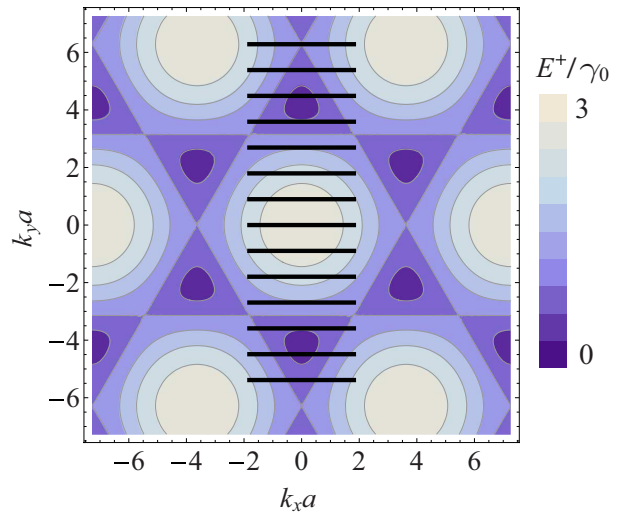}
  \caption{%
    (color online) Graphene Band structure and subbands for (7,0)
    nanotube bands. The transitions between bands allowed with the electric field perpendicular to the nanotube can occur between neighboring subbands.}
  \label{fig2}
\end{figure}

\begin{figure}[t]
  \centering\includegraphics[width=8cm]{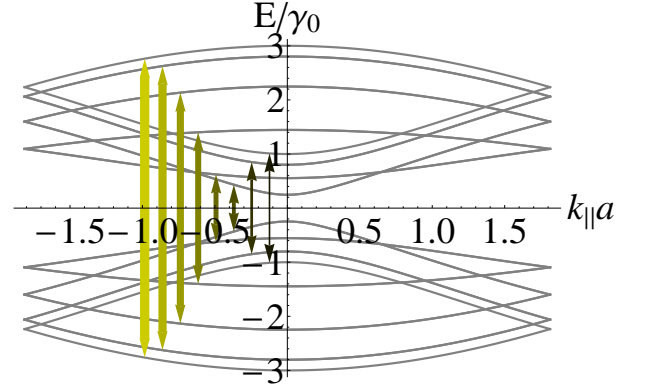}
  \caption{%
    (color online) The dispersion relation for a (7,0) nanotube.
    The possible transitions caused by light linearly polarized parallel to the nanotube
    are shown by arrows. Here the bandgap is $1.43$eV, when $\gamma_0$ is taken as $2.89$eV.}
  \label{fig3}
\end{figure}

The well-established tight-binding model assumes that the electrons
are tightly bound to the individual carbon atoms and the localized
atomic orbitals are used as a basis for expanding the wavefunction. We
consider orbitals that contribute to states that lie within an optical
range of energies around the Fermi level. These are the states that
give the main contributions to the optical properties
of the nanotube. Every carbon atom has four valence orbitals
(2s, 2$p_x$, 2$p_y$ and 2$p_z$) that could lie in this energy range. For
2D graphene the (s, $p_x$, $p_y$) orbitals combine to form hybridized
$sp^2$ orbitals. These give the strong covalent bonds; primarily
responsible for the binding energy and elastic properties of the
nanotubes. In the tight-binding model they result in $\sigma$ and
$\sigma^*$ bands. However, their energy levels are far away from the
Fermi level and hence they do not play a key role in the optical properties that we are interested in. That role is played by delocalized $\pi$ and $\pi^*$ bands that are formed from the $p_z$
orbitals~\cite{RevModPhys.79.677}. Hence, we can ignore the $\sigma$
electrons and restrict the tight-binding model to the $\pi$ electrons.
The Hamiltonian for this system is
\begin{equation}
  \hat{H}_0=-\gamma_0 \sum_{ij}(\hat{\alpha}^\dagger_i \hat{\beta}_j + h.c.),
  \label{eq1}
\end{equation}
where $-\gamma_0$ is the hopping amplitude and $ij$ refers to nearest
neighbors. Here, $\hat{\alpha}^\dagger_i$ and $\hat{\beta}^\dagger_j$
are the creation operators for electrons in sublattice A and B,
respectively. In this Hamiltonian we have removed the constant energy
contribution that corresponds to the Fermi level. We expand the wavefunction in terms of $p_z$ orbitals
at every atom site and split this expression into two parts; one for
each sublattice. The wavefunction for each state is then
\begin{align}
  \Psi(\mathbf{k},\mathbf{r})&=\sum_{\mathbf{r}_A} C_A(\mathbf{r}_A,\mathbf{k})p_z(\mathbf{r}-\mathbf{r}_A) \\
  &+\sum_{\mathbf{r}_B}C_B(\mathbf{r}_B,\mathbf{k})
  p_z(\mathbf{r}-\mathbf{r}_B)
  \label{eq2}
\end{align}
with $\mathbf{r}_A$,$\mathbf{r}_B$ labeling the atom locations in sublattice A and B, respectively. The
individual $p_z$ orbitals are given by the normalized wavefunctions
$p_z(\mathbf{r})$ and each one has a coefficient, represented
with $C_A$ and $C_B$. By using translational symmetry we can represent
this as
\begin{equation}
  \Psi(\mathbf{k},\mathbf{r})= c_A(\mathbf{k})\tilde{p}_z^A (\mathbf{k},\mathbf{r})+c_B(\mathbf{k})\tilde{p}_z^B (\mathbf{k},\mathbf{r})
  \label{eq3}
\end{equation}
where the Bloch functions, $\tilde{p}_z^A$ and $\tilde{p}_z^B$, are
\begin{align}
  \tilde{p}_z^A (\mathbf{k},\mathbf{r}) &=\frac{1}{\sqrt{N_{cells}}}\sum_{\mathbf{r}_A}\e^{\ii\mathbf{k}.\mathbf{r}_A}p_z(\mathbf{r}-\mathbf{r}_A) \\
  \tilde{p}_z^B (\mathbf{k},\mathbf{r})
  &=\frac{1}{\sqrt{N_{cells}}}\sum_{\mathbf{r}_B}\e^{\ii\mathbf{k}.\mathbf{r}_B}p_z(\mathbf{r}-\mathbf{r}_B).
  \label{eq4}
\end{align}
Here $N_{cells}$ is the number of unit cells in the graphene sheet. These bloch functions have the coefficients $c_A$ and $c_B$. The
states are labeled by their crystal momentum vector
$\mathbf{k}$. We now solve the time-independent
single-particle Schr\"odinger equation
\begin{equation}
  \hat{H}_0 \Psi(\mathbf{k},\mathbf{r}) = E(\mathbf{k}) \Psi(\mathbf{k},\mathbf{r}).
  \label{eq5}
\end{equation}
We define the quantity
\begin{equation}
\phi_{\mathbf{k}} = \sum_{\mathbf{q}} \e^{\ii \mathbf{k}.\mathbf{q}},
  \label{eq6}
\end{equation}
where $\mathbf{q}$ are vectors from an atom in the A sublattice to its
neighboring atoms in the B lattice.  This gives us the following
quantities
\begin{align}
  H_{AA}=H_{BB}&=\bra{\tilde{p}_z^A} \hat{H}_0 \ket{\tilde{p}_z^A} =\bra{\tilde{p}_z^B} \hat{H}_0 \ket{\tilde{p}_z^B}=0 \notag \\
  H_{AB}=H_{BA}^* &=\bra{\tilde{p}_z^A} \hat{H}_0 \ket{\tilde{p}_z^B}=-\gamma_0 \phi_{\mathbf{k}} \notag \\
  S_{AB}=S_{BA}^* &=\langle \tilde{p}_z^A | \tilde{p}_z^B \rangle = u \phi_{\mathbf{k}}.
  \label{eq7}
\end{align}
Now the variational Schr\"odinger equation in matrix form is
\begin{equation}
    \begin{pmatrix}
      H_{AA} & H_{AB} \\
      H_{BA} & H_{BB}
    \end{pmatrix}
    \begin{pmatrix}
      c_A \\
      c_B
    \end{pmatrix}
 =E(\mathbf{k})
    \begin{pmatrix}
      1 & S_{AB} \\
      S_{BA} & 1
    \end{pmatrix}
    \begin{pmatrix}
      c_A \\
      c_B
    \end{pmatrix}.
  \label{eq8}
\end{equation}
The above matrix equation can be solved to give the energy of each state as
\begin{align}
  E^\pm(\mathbf{k}) &= \frac{\pm \gamma_0 \left|\phi_{\mathbf{k}}\right|}{1 \mp u  \left|\phi_{\mathbf{k}}\right|},
   \label{eq9}
\end{align}
where
\begin{align}
  \left|\phi_{\mathbf{k}}\right| &= \Bigg( 1+ 4\cos \left(\frac{k_x
      a\sqrt{3}}{2}\right)\cos \left(\frac{k_y a}{2}\right) \nonumber\\
&\qquad\qquad\qquad\qquad\qquad + 4\cos^2
    \left(\frac{k_y a}{2}\right) \Bigg)^{\frac{1}{2}}.
  \label{eq9b}
\end{align}
 A typical value for $\gamma_0$ is $2.89$eV such that the tight-binding model corresponds with experiments~\cite{PhysRevB.67.165402,PhysRevB.61.2981,saito1998physical}. We will keep $u$ in the equations but for all plots and numerical calculations we assume that $u=0$, i.e. there is no orbital overlap. In Fig.~\ref{fig2} this 2D
dispersion relation is plotted as a contour. In Eq.~(\ref{eq9}) the
signs refer to the two relevant bands, the conduction band and
the valence band. The coefficients are found to be
\begin{align}
  c^v_A(\mathbf{k})&=\sqrt{\frac{\phi_{\mathbf{k}}}{2\left|\phi_{\mathbf{k}}\right|(1+u\left|\phi_{\mathbf{k}}\right|)}}, \\
  c^v_B(\mathbf{k})&=\sqrt{\frac{\phi_{\mathbf{k}}^*}{2\left|\phi_{\mathbf{k}}\right|(1+u\left|\phi_{\mathbf{k}}\right|)}}, \\
  c^c_A(\mathbf{k})&=-\sqrt{\frac{\phi_{\mathbf{k}}}{2\left|\phi_{\mathbf{k}}\right|(1-u\left|\phi_{\mathbf{k}}\right|)}}, \\
  c^c_B(\mathbf{k})&=\sqrt{\frac{\phi_{\mathbf{k}}^*}{2\left|\phi_{\mathbf{k}}\right|(1-u\left|\phi_{\mathbf{k}}\right|)}}.
  \label{eq10}
\end{align}
Now we know the relevant band structure of graphene and their
wavefunctions, we need to obtain the energy states of the nanotubes.
To do this we use the zone-folding approximation. This assumes the
nanotube bands are the same as graphene but with limited $\mathbf{k}$,
due to the 1D nature of a carbon nanotube. There are a variety of ways
available to wrap the sheet up into a nanotube, each of which result
in very different properties. The nanotubes are characterized by a
vector in the graphene plane that corresponds to the circumference of
the nanotube and is called the chiral vector $\mathbf{C}=n_1
\mathbf{a}_1+n_2 \mathbf{a}_2$ $(0\leq \vert n_2 \vert \leq
n_1)$ (see Fig.~\ref{fig1}). It gives the relative position of two
graphene atoms that become `identical' when the graphene is rolled
into a nanotube. We will use these parameters in the standard form $(n_1,n_2)$ to label each type of nanotube. This immediately defines some basic
geometric properties such as the tube's circumference and radius
$R_t=a \sqrt{n_1^2+n_1 n_2 +n_2^2}/2\pi$.
We also define the translational vector, perpendicular to
$\mathbf{C}$, that corresponds to the direction along the tube,
$\mathbf{T}=t_1 \mathbf{a}_1+t_2 \mathbf{a}_2$.  Using the greatest
common divisor (gcd), we define $t_1=(2n_2 +n_1)/N_R$,
$t_2=-(2n_1 +n_2)/N_R$ and $N_R=\gcd(2n_1 +n_2,2n_2 +n_1)$. The
two vectors, $\mathbf{C}$ and $\mathbf{T}$, define the unit cell of
the nanotube. Within each nanotube unit cell there are
$N_G=2(n_1^2+n_1 n_2 +n_2^2)/N_R$ graphene unit cells and, hence,
$N_C=2N_G$ carbon atoms. In a nanotube of length $L$ there are $N_L=L/\left|\mathbf{T}\right|$ nanotube unit cells. For the nanotube's reciprocal lattice we define $\mathbf{K}_1=(t_1 \mathbf{b}_2-t_2 \mathbf{b}_1)/N_G$ and
$\mathbf{K}_2=(n_2 \mathbf{b}_1-n_1 \mathbf{b}_2)/N_G$ such that
$\mathbf{K}_1.\mathbf{T}=\mathbf{K}_2.\mathbf{C}=0$ and
$\mathbf{K}_1.\mathbf{C}=\mathbf{K}_2.\mathbf{T}=2\pi$. These give the
allowed vectors in the SWCNT's Brillouin zone to be a set of $N_G$ 1D
`cutting lines' with values
\begin{equation}
  \mathbf{k}=\mu \mathbf{K}_1 + k_{||} \frac{\mathbf{K}_2}{\left| \mathbf{K}_2  \right|},
  \label{eq11}
\end{equation}
with $\mu = -N_G/2+1,\ldots,0,\ldots,N_G/2$ and $-\pi/\left|\mathbf{T}\right|
\leq k_{||} < \pi/\left|\mathbf{T}\right|$. It is the periodic
boundary condition along the circumferential direction of the tube
that causes the wave vector to become quantized and each discrete
cutting line is labeled by the azimuthal quantum number $\mu$.
For short nanotubes the wave vectors are also quantized along the nanotube's length causing discrete energy levels to be formed~\cite{PhysRevB.79.045418}. These discrete values have $k_{||}=2\pi j/L-\pi/\left|\mathbf{T}\right|$, for an integer $j=1,\ldots,N_L$. Local
effects also occur in short nanotubes, such as a sharp spike in the
density of states (DOS), caused by defects at the caps. Such effects
will be ignored here. Typically, the nanotube is assumed to be of infinite length, allowing
continuous values of the wave vector along the nanotube axis. This
causes possible wave vectors to lie in `subbands'. The subbands can cut through the fermi points of
graphene causing the tube to become metallic. This can be shown to be
the case for nanotubes of the type $(n,m)$, where $n-m$ is a multiple
of three. If this is not the case there is a nonzero band gap and the
nanotube is semiconducting. Here we consider `zigzag' semiconducting
nanotubes of the form $(n,0)$, with $n$ not a multiple of three. The
discrete wave vectors are then
\begin{equation}
  \left|k_{\bot}\right|=\frac{2\pi \mu}{\left|\mathbf{C}\right|}
  \label{eq12}
\end{equation}
with $\mu=-(n-1),...,0,1,2,...,n$ and
\begin{equation}
  \left|k_{||}\right|<\frac{\pi}{\left|\mathbf{T}\right|}.
  \label{eq13}
\end{equation}
The momentum vectors associated with these subbands are highlighted in
Fig.~\ref{fig2} for a (7,0) nanotube. For zigzag nanotubes $k_{\bot}$ corresponds to $k_y$ and $k_{||}$ corresponds to $k_x$. It should be noted that subbands
$\mu$ and $-\mu$ both have the same energy and this degeneracy is
referred to as the `valley' degeneracy. Combined with the two electron
spins this gives a degeneracy of four for each energy value, except for $\mu=0$ and $\mu=n$ that only have spin degeneracy. The
energy of the subbands is
\begin{equation}
  E^{\pm}_{NT}(k_{||},\mu)=E^{\pm}(k_{||}\frac{\mathbf{K}_2}{|\mathbf{K}_2|}+\mu \mathbf{K}_1)
  \label{eq14}
\end{equation}
These are plotted for a (7,0) nanotube in Fig.~\ref{fig3}, which has a
bandgap of $1.43$eV.

\section{The Optical Absorption of Carbon Nanotubes \label{sec3}}

The Hamiltonian of a nanotube interacting with an electromagnetic
field is $\hat{H}=\hat{H_0}+\hat{H_F}+\hat{H_I}$, with
\begin{equation}
  \hat{H}_I = \frac{e}{m_e} \hat{\mathbf{A}}.\hat{\mathbf{p}},
  \label{eq15a}
\end{equation}
being the interaction term and $\hat{H_F}$ representing the field
Hamiltonian. Here, we define $e$ as the magnitude of the electron charge and
are using SI units. Each field mode is characterized by its angular
frequency $\omega$ and further parameters, which define the mode's
polarization and propagation direction. The field vector potential
operator is
\begin{equation}
  \hat{\mathbf{A}} = \int_{0}^{\infty}d\omega (\hat{\mathbf{A}}_{\omega}^{+} \e^{-\ii \omega t} + \hat{\mathbf{A}}_{\omega}^{-} \e^{\ii \omega t}).
  \label{eq16}
\end{equation}
We will consider the initial and final state of field to be a coherent
monochromatic state $\ket{\alpha_{\omega_0}}$, with an angular
frequency of $\omega_0$ and a mean photon flux of $F$ photons per unit
time. This state satisfies the equation $\hat{a}_{\omega}
\ket{\alpha_{\omega_0}}=\alpha \ket{\alpha_{\omega_0}}$, with $\alpha
=\sqrt{2\pi F} \delta(\omega-\omega_0)$ and $\hat{a}_{\omega}$ being
the field mode's destruction operator~\cite{PhysRevA.42.4102}. This
allows us to give
\begin{align}
  \mathbf{A} &= \bra{\alpha_{\omega_0}} \hat{\mathbf{A}} \ket{\alpha_{\omega_0}} \\
  &= \mathbf{A}^{+} \e^{-\ii \omega_0 t} + \mathbf{A}^{-} \e^{\ii
    \omega_0 t}.
  \label{eq16b}
\end{align}
%
Using time-dependent perturbation theory we find that, after time $t$, the initial state of the nanotube and field, $\ket{\Psi^v}\ket{\alpha_{\omega_0}}$, is in the state $\ket{\Psi'^c}\ket{\alpha_{\omega_0}}$ with probability
\begin{equation}
  P = t \frac{2\pi}{\hbar}\left| \bra{\Psi'^c} (\frac{e}{m_e} \mathbf{A}^+.\hat{\mathbf{p}}) \ket{\Psi^v} \right|^2 \delta(E'-E-\hbar\omega_0),
  \label{eq17}
\end{equation}
which is Fermi's Golden Rule. The transition rate for each electron in
the state with energy $E$ to each state with energy $E'$ can be
expressed as
\begin{equation}
  w = \frac{2\pi}{\hbar}\left| \frac{e}{m_e} \ii \hbar G \right|^2 \delta(E'-E-\hbar\omega_0).
  \label{eq18}
\end{equation}
To calculate the optical absorption of a carbon nanotube of length $L$, the interaction term, $\ii\hbar G=\ii\hbar
\bra{\Psi'^c} \mathbf{A}^+.\nabla\ket{\Psi^v}$, needs to be found with spatially changing field. Here we are assuming that the state is coherent over the entire length of the nanotube. To calculate $G$ we define
\begin{align}
  \mathbf{v}^A(\mathbf{k}) &=  \sum_{\mathbf{q}} \e^{\ii \mathbf{k}.\mathbf{q}} \mathbf{q},  \\
  \mathbf{v}^B(\mathbf{k}) &=  - \sum_{\mathbf{q}} \e^{-\ii \mathbf{k}.\mathbf{q}} \mathbf{q},
  \label{eq28}
\end{align}
with $\mathbf{q}$ summing over the three vectors pointing from an atom in the A sublattice to its neighboring three B lattice atoms. We will furthermore use the matrix element, $M=\bra{p_z(\mathbf{r})}\nabla_z\ket{p_z(\mathbf{r}-\mathbf{q}_z}$, with $\mathbf{q}_z$ being the vector between two neighboring atoms such that the z-axis is aligned along $\mathbf{q}_z$. The value we will later use for this is given by~\cite{PhysRevB.74.155434}
\begin{equation}
   M = 2 a \gamma_0 m_e / \hbar^2 \sqrt{3}.
  \label{eq28b}
\end{equation}
Each of the unit cells in the nanotube extends over a distance of $|\mathbf{T}|\approx 0.43\,$nm along the nanotube and approximately a nanometre across. This is much smaller than the light's wavelength and spatial variations. Therefore, we can assume that the electromagnetic field is constant across each of the nanotube's unit cells. There are $N_L$ of these unit cells along the nanotube's length and in each one, labeled by an integer $l$, the field is given by $\mathbf{A}^+_l=\mathbf{A}^+(l|\mathbf{T}|-L/2)$.

An expression for $G$ can then be calculated and simplified into the form
(see Appendix~\ref{app1})
\begin{align}
  G &= \frac{1}{N_L} \mathbf{D}(\mathbf{k}',\mathbf{k}).\left[ \sum_{l=1}^{N_L} \e^{\ii (al\sqrt{3}-L/2) (k_{||}-k_{||}')} \mathbf{A}^+_l \right] \\
  &\approx \frac{1}{L} \mathbf{D}(\mathbf{k}',\mathbf{k}).\left[
    \int_{l=-L/2}^{L/2} dl \e^{\ii l (k_{||}-k_{||}')}
    \mathbf{A}^+(l) \right],
  \label{eq19}
\end{align}
where
\begin{align}
  D_z &= \frac{M\sqrt{3}}{2 a n} \sum_{j=1}^{n} (c_A^{c*}(\mathbf{k}') c_B^{v}(\mathbf{k}) \e^{-\ii j a (k_{\bot}'-k_{\bot})} \notag \\
  & (1+\e^{-\ii a (\mathbf{k}'-\mathbf{k}).(\sqrt{3}/2,1/2)}) v_z^A (\mathbf{k})  \notag \\
  & - c_B^{c*}(\mathbf{k}') c_A^{v}(\mathbf{k}) \e^{-\ii j a (k_{\bot}'-k_{\bot})} \e^{-i a (k_{||}'-k_{||})/\sqrt{3}} \notag \\
  & (1+\e^{-\ii a (\mathbf{k}'-\mathbf{k}).(\sqrt{3}/2,1/2)})
  v_z^A (\mathbf{k})^*),
  \label{eq20}
\end{align}
and $D_{x,y}$ are given in Appendix~\ref{app1}.

\begin{figure}[t]
  \centering\includegraphics[width=8cm]{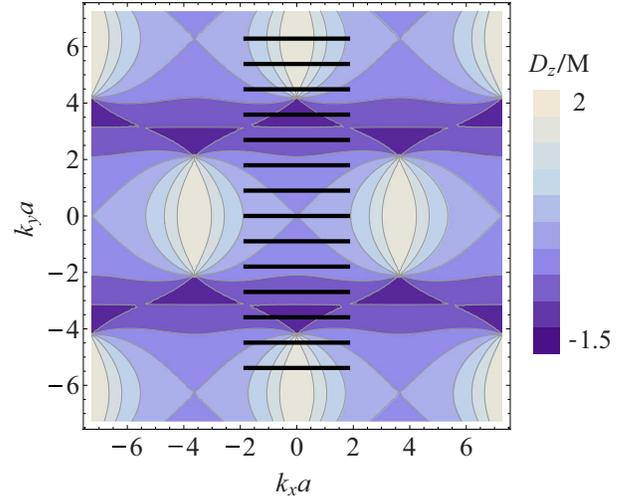}
  \caption{%
    (color online) A contour plot of
    $D_z(\mathbf{k},\mathbf{k})$ for graphene is shown
    together with the nanotube's subband lines. It is given in terms
    of the constant $M$ from Eq.~(\ref{eq28b}).}
  \label{fig4}
\end{figure}

\begin{figure}[t]
  \centering\includegraphics[width=8cm]{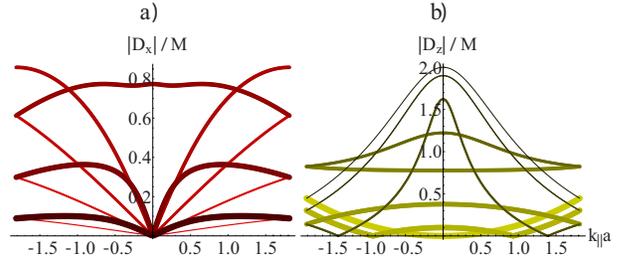}
  \caption{%
    (color online) a) $D_x$ values for the different
    transitions that can occur when the electric field is
    perpendicular to the nanotube. $M$ is given by Eq.~(\ref{eq28b}). b)
    $D_z$ for transitions allowed with an electric field
    parallel to the nanotube.}
  \label{fig5}
\end{figure}

This result coincides with that of Ref.~\cite{Jiang20043169} when
$\mathbf{A}^+_l$ is the same for all $l$. In Eq.~(\ref{eq19}) the
$\mathbf{D}$ gives the selection rules for possible transitions between bands $\mu'$ and $\mu$.
In particular $D_z(\mathbf{k}',\mathbf{k})$ is negligible if $\mu' \neq \mu$ and
for the other components of $\mathbf{D}(\mathbf{k}',\mathbf{k})$ to contribute we require $\mu' =\mu \pm 1$. For a uniform field across the nanotube these give the possible transitions for a field polarized parallel and perpendicular to the nanotube, respectively. In Fig.~\ref{fig4} we have plotted $D_z(\mathbf{k},\mathbf{k})$, and in Fig.~\ref{fig5} $D_{x,z}$ is
plotted. These expressions correspond to direct transitions, i.e with $k_{||}' = k_{||}$, which is an approximation of momentum conservation and will be discussed later in this section.

Although the values of $D_x$ and $D_y$ show a
transition, the induced local field creates a depolarization
effect~\cite{PhysRevB.57.9301,Ajiki1994349,PhysRevLett.93.037404} that reduces
$D_x$ and $D_y$ to give a negligible contribution to
the absorption. This allows us to focus on the $D_z$ term and
simplify $G$ to
\begin{equation}
  G \approx \frac{1}{L} D_z(\mathbf{k}',\mathbf{k})\left[ \oint \mathbf{A}^+.\mathbf{dr} \e^{\ii s(k_{||}-k_{||}')}  \right],
  \label{eq21}
\end{equation}
with $s$ denoting the length along the nanotube. This also restricts the transitions to those with $\mu'=\mu$.

It is the line integral in Eq.~(\ref{eq21}) that is responsible for momentum conservation in the system. The photon momentum is much smaller than the crystal momentum and typically only direct transitions are considered, i.e. $k_{||}' \approx k_{||}$. Here we will make this assumption, however the change in momentum can not be completely neglected since any change can make a major difference to the line integral in Eq.~(\ref{eq21}). This is especially true when the field oscillates along the nanotube. The energy of a direct transition is given by $E_g(\mathbf{k})=E^+(\mathbf{k})-E^-(\mathbf{k})$. Since $D_z(\mathbf{k}',\mathbf{k}) \approx D_z(\mathbf{k},\mathbf{k})$, when $k_{||}'\approx k_{||}$, we will make this substitution and further simplify $D_z(\mathbf{k},\mathbf{k})=D_z(\mathbf{k})$ to give
\begin{equation}
  D_z(\mathbf{k}) =  \frac{-M\sqrt{3}}{a} Re\left( v_z^A (\mathbf{k}) \frac{\phi_{\mathbf{k}}^*}{\left|\phi_{\mathbf{k}}\right|\sqrt{1-u^2\left|\phi_{\mathbf{k}}\right|^2}}\right).
  \label{eq22}
\end{equation}
We define $A^+_{||}(s)ds=\mathbf{A}^+.\mathbf{dr}$ to be the field potential along the nanotube and use the discrete momentum values, $k_{||}=2\pi j/L$ and $k_{||}'=2\pi j'/L$, with integers $j$ and $j'$. The line integral can then be expressed as
\begin{align}
  S(k_{||}-k_{||}') &= (1/L)\oint \mathbf{A}^+.\mathbf{dr} \e^{\ii s(k_{||}-k_{||}')} \\
  &= (1/L)\int_{-L/2}^{L/2} ds A^+_{||}(s) \e^{\ii s(k_{||}-k_{||}')} \\
  &= (1/L)\int_{-L/2}^{L/2} ds A^+_{||}(s) \e^{\ii 2\pi s(j-j')/L}.
  \label{eq22b}
\end{align}
This expression is simply the coefficient in the Fourier series for
$A^+_{||}(s)$. Since the photon momentum is very small in comparison to the
crystal momentum the only relevant coefficients will have very small
values of $j'-j$ relative to $N_L$. Every electron transition in the
nanotube then needs to be considered to calculate the overall
absorption rate. This leads to a length dependence on the absorption.
We initially consider discrete states and corresponding $k_{||}$
values. The transition rate given by Eq.~(\ref{eq18}) is summed over all
possible initial and final states to give
\begin{align}
  w_{L} &\approx \sum_{d_i}\sum_{\mu=-n+1}^n  \sum_{k_{||}} \sum_{k_{||}'} \frac{2\pi \hbar e^2}{m_e^2} \left|D_z(\mathbf{k})\right|^2 \notag \\
  & \qquad\qquad \left|S(k_{||}-k_{||}')\right|^2
  \delta(E_g(\mathbf{k})-\hbar\omega_0).
  \label{eq23}
\end{align}
In this equation $d_i$ refer to the degeneracy of the initial state.
For any value of $k_{||}$ the sum over $k_{||}'$ causes
$k_{||}-k_{||}'$ to take all of the low values that are relevant. This
sum is also independent of $k_{||}$ and allows us to define
$S=\sum_{j} \left|S(2\pi j/L)\right|^2$, which can be rewritten using
Parseval's theorem to be
\begin{equation}
  S=(1/L)\int_{-L/2}^{L/2} ds \left| A^+_{||}(s)\right|^2 .
  \label{eq23c}
\end{equation}
The total absorption rate is then
\begin{align}
  w_{L} &\approx \sum_{d_i}\sum_{\mu=-n+1}^n  \sum_{k_{||}} \frac{2\pi \hbar e^2}{m_e^2} \left|D_z(\mathbf{k})\right|^2 \notag \\
  & S \delta(E_g(\mathbf{k})-\hbar\omega_0) \\
  &\approx U(\omega_0) \int_{-L/2}^{L/2} ds \left| A^+_{||}(s)\right|^2 ,
\label{eq23b}
\end{align}
where we define
\begin{equation}
  U(\omega_0)=\sum_{d_i}\sum_{\mu=-n+1}^n  \int dk_{||} \frac{\hbar e^2}{m_e^2} \left|D_z(\mathbf{k})\right|^2 \delta(E_g(\mathbf{k})-\hbar\omega_0).
  \label{eq23d}
\end{equation}
The field was defined to be a coherent state with a photon flux given by $F$ photons per unit time. We divide the transition rate, given in Eq.~(\ref{eq23b}), by this flux to obtain an estimate for the probability of one photon exciting a single electron.
This expression gives a probability that increases linearly with nanotube length. This is certainly suitable up to the coherence length, $L_c$, however not for long nanotubes. So far we have considered the length of the nanotube to be smaller than the coherence length. For long nanotubes we can consider the whole nanotube to be composed of coherent segments. This leads to an
exponential increase in the absorption with the nanotube length. Here,
to calculate this quantity we find the probability of not exciting any
electrons, which is the product of $(1-w_{L_c}/F)\approx e^{-w_{L_c}/F}$ for each segment. Hence, the probability of exciting a single electron, in a nanotube of length $L$, with each photon can be estimated by the expression
\begin{equation}
  \eta = 1-\exp \left( \frac{-U(\omega_0)}{F} \int_{-L/2}^{L/2} ds \left| A^+_{||}(s)\right|^2 \right).
  \label{eq23e}
\end{equation}
Note that this expression is actually independent of the coherence length.

Broadening effects can be included in this by substituting the Dirac delta function, from Eq.~(\ref{eq23d}) with a Lorentzian
function,
\begin{equation}
  \delta(E_g-\hbar\omega) \rightarrow \frac{\Gamma}{\pi((E_g-\hbar\omega)^2+\Gamma^2)},
  \label{eq24}
\end{equation}
that has a broadening parameter, $\Gamma$. This parameter can include the broadening due to multiple effects, including the electronic state's decay. In carbon nanotubes the state decay occurs on a picosecond time scale~\cite{PhysRevLett.90.057404}. If we take a range of $0.1$ps to $2$ps the required broadening ranges from $\Gamma = 0.01$eV to $\Gamma = 0.001$eV. In order to compare our results with previous work~\cite{PhysRevB.74.155434,PhysRevB.62.13153} we will choose in the following to use a parameter of $\Gamma = 0.01$eV.

So far we have assumed the light to be in a pure state consisting of one specific wavelength. We expect to have a range of wavelengths present and to deal with this we assume the light is in a probabilistic mixture of coherent beams, each with a photon flux $F$. These are weighted by a lineshape $g(\omega)$, satisfying $\int d\omega g(\omega)=1$. The light's state is then $\int d\omega g(\omega) \ket{\alpha_{\omega}}\bra{\alpha_{\omega}}$ and the expected absorption probability is
 \begin{equation}
  \overline{\eta} =\int d\omega g(\omega) \eta.
  \label{eq24b}
\end{equation}
In the following we take $g$ to be a uniform lineshape between two energy values. This
is equivalent to taking $\overline{\eta}$ to be the average transition
probability over a range of energies.

\section{Optical Nanofiber Photon Absorption into a Carbon Nanotube \label{sec4}}
We now extend the calculation for the absorption around optical fibers
and particularly nanofibers~\cite{ISI:000187342000051,Tong20081240}. A
review of these subwavelength diameter waveguides can be found in
Refs.~\cite{Brambilla:04,springerlink:10.1007/s12200-009-0073-1,ISI:000220594800008}. They are made of a silica core and
have diameters as small as $50\,$nm. For fibers of this size a high
proportion of the light field exists outside of the fiber's core. This
means the field is easily accessible and we consider positioning the
carbon nanotube near the nanofiber. The use of fibers allows the
interaction to be enhanced due to the transverse confinement of the
field. Altering the nanofibers properties also allows us to tailor the
field. Fibers with a smaller diameter have a larger evanescent field
but also suffer from higher losses. We will consider a cylindrical
nanofiber core with a radius of $R$ and cladding provided by the
vacuum, with refractive index $n_2 = 1$. The refractive index of the
silica core is $n_1 = 1.45$ and the material absorption of the silica
is negligible over the short distances being considered. Silica core
fibers with subwavelength diameter are single mode fibers, i.e. the
only mode present is the HE fundamental mode
(see Ref.~\cite{ISI:000220594800008} for a general single mode condition).
In the following we will adopt a scheme used
in Refs.~\cite{PhysRevA.72.063815,PhysRevA.42.4102} to quantize the field.
The field potential operator for the nanofiber's fundamental mode is
then
\begin{equation}
  \hat{\mathbf{A}}_{\omega}^{+} = \sum_{f p} \sqrt{\frac{\hbar \beta '}{4 \pi \omega \epsilon_0 A}}\hat a_{m} \mathbf{e}^{m}(r,\varphi) \e^{\ii(f\beta z +p\varphi)}.
  \label{eq25}
\end{equation}
\begin{figure}[t]
  \centering\includegraphics[width=8cm]{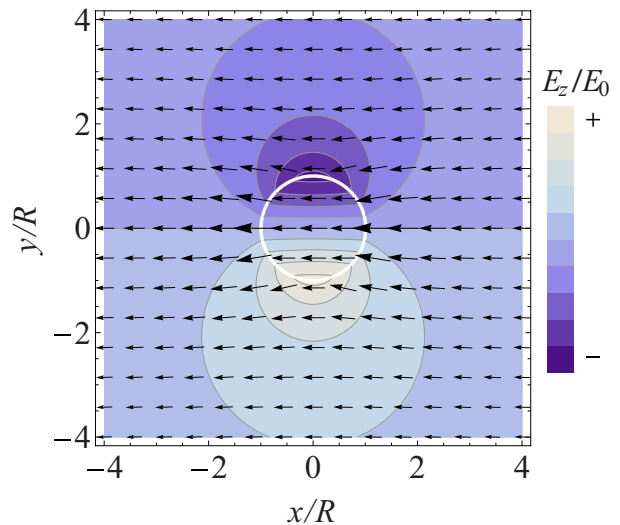}
  \caption{%
    (Color online) The electric field, $\mathbf{E}$, of the circularly polarized
    HE mode taken for a constant $z$, to give a cross-section of the
    fiber. The contour gives $E_z$ and the arrows represent the
    $x$ and $y$ components. The longer the arrow the stronger the
    field. The field has been divided by the constant $E_0=\sqrt{F \omega_0 \hbar \beta '/(2 \epsilon_0 A)}$. In this particular case we took the fiber diameter to be $250$nm and the light's wavelength as $868$nm.}
  \label{fig6}
\end{figure}
This is given in terms of cylindrical coordinates, with $z$ being the
coordinate along the fiber and $\varphi$ the azimuthal angle. The
light's angular frequency is $\omega$. Its propagation direction is
labeled with $f=\pm 1$ and $\beta$ refers to the longitudinal
propagation constant. We find the value of $\beta$ by numerically
solving the fiber eigenvalue equation (see Eq.~(\ref{eq32}) in
Appendix~\ref{app2}).  The derivative in Eq.~(\ref{eq25}), $\beta'$,
is taken with respect to $\omega$, and $\hat{a}_{m}$ are the photon
annihilation operators, with $m = (\omega ,f ,p)$ characterizing the
separate modes. Furthermore, $\mathrm{\mathbf{e}}^{m}$ are the
electric field profiles of the guided mode that can be found by
solving Maxwell's
equations~\cite{snyder1983optical,PhysRevA.72.063815} and $A$ gives a
normalization factor. The expressions for the mode profiles and $A$
are given in Appendix~\ref{app2}. The polarization can be right or
left circular labeled by $p=\pm 1$. For a single mode of monochromatic
coherent light with $m=(\omega_0 ,f ,p)$ we have
\begin{equation}
  \mathbf{A}^{+} = \sqrt{\frac{F \hbar \beta '}{2 \omega_0 \epsilon_0 A}} \mathbf{e}^{m}(r,\varphi) \e^{\ii(f\beta z +p\varphi)}.
  \label{eq25ii}
\end{equation}
Fig.~\ref{fig6} provides a slice of the classical field at one instant
in time. This field can be seen to extend far away from the nanofiber
and vary dramatically with position.

\begin{figure}[t]
  \centering\includegraphics[width=8cm]{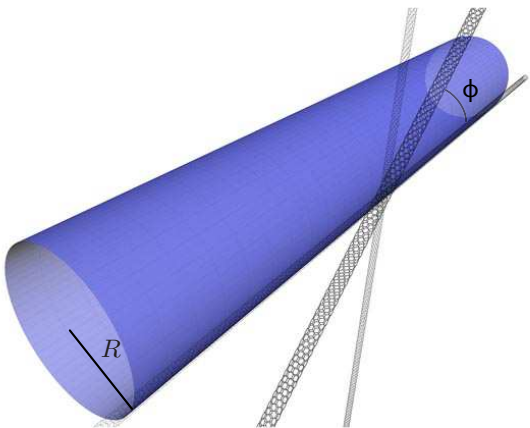}
  \caption{%
    (color online) Possible orientations of a straight
    nanotube relative to a fiber of radius $R$. $\phi$ labels the angle between the nanofiber and nanotube.}
  \label{fig7}
\end{figure}

This contrasts with the simpler case, that has previously been
studied, of a plane linearly polarized light beam that is
given by
\begin{equation}
  \hat{\mathbf{A}}_{\omega}^{+} = \sqrt{\frac{\hbar}{4 \pi \omega \epsilon_0 c A'}}\hat{a}_{\omega} \mathbf{e}_z,
  \label{eq25b}
\end{equation}
across the whole nanotube, where the beam has a finite cross-sectional
area of $A'$. This gives
\begin{equation}
  \mathbf{A}^{+} = \sqrt{\frac{F \hbar}{2 \omega_0 \epsilon_0 c A'}} \mathbf{e}_z.
  \label{eq25bii}
\end{equation}
For fibers larger than $100\,$nm in diameter the photon losses are
small and can be ignored over short distances. In our calculations we
will use nanofibers of diameter $250\,$nm. Furthermore, we will focus
on the forward propagation and right polarized guided modes, i.e.
$f=p=+1$. All other modes are related to our results by symmetry.
The value of $G$ is then highly
dependent on the way the nanotube is orientated relative to the nanofiber and can be calculated to be
\begin{align}
  G &= \sqrt{\frac{F \hbar \beta '}{2 \omega_0 \epsilon_0 A}} \notag \\
  & \bra{\Psi^c (\mathbf{k}')} \mathbf{e}^{m}\exp^{\ii(f\beta z
    +p\varphi)}.\nabla \ket{\Psi^v (\mathbf{k})}.
  \label{eq26}
\end{align}
Since the field strength drops off exponentially the highest value for
the coupling will be achieved by having the nanotube as close as
possible to the fiber. In our examples, the distance between the nanotube's center and the surface of the nanofiber
is chosen to be $1.25\,$nm. The nanotubes we consider always have a radius less than $1\,$nm so this distance avoids any contact.
There are two orientations we will consider. The first is that of a
straight nanotube, of length $L$, oriented at an angle $\phi$ relative
to the nanofiber which includes parallel and perpendicular
orientations as illustrated in Fig.~\ref{fig7}.

\begin{figure}[t]
  \centering\includegraphics[width=8cm]{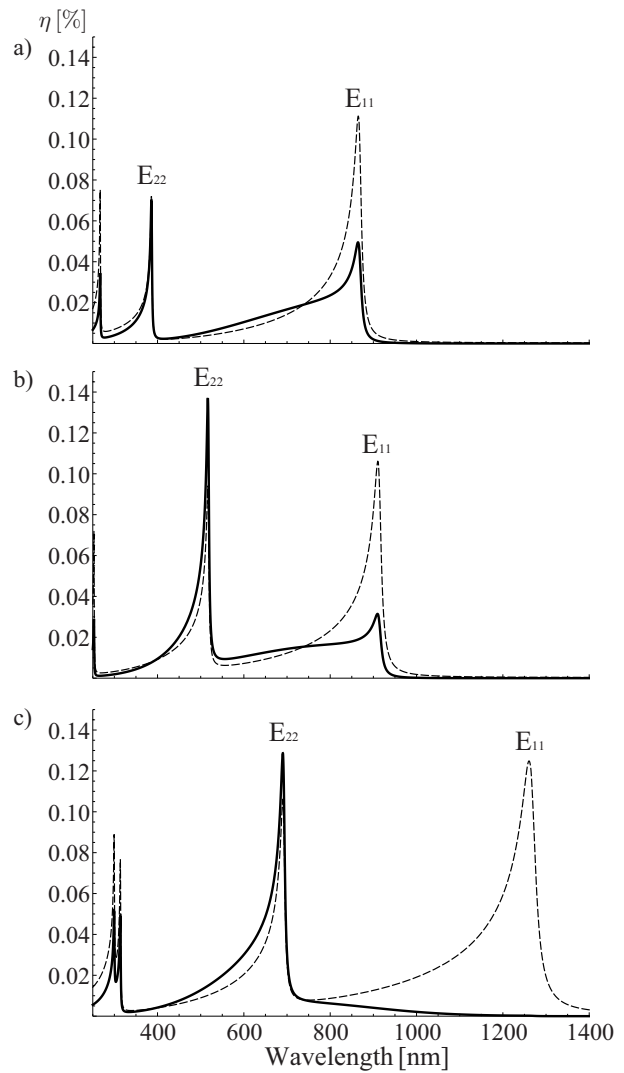}
  \caption{%
    Photon absorption probabilities, $\eta$, for a
    $2\,\mu$m long nanotube, perpendicular to the fiber, at different
    photon energies. The nanotubes considered are a) (7,0) b) (8,0) and c) (11,0). In each case the solid lines refer to the absorption of circularly polarized light guided by the nanofiber and the dashed lines represent the absorption for a plane coherent light beam (without a fiber) that is linearly polarized along the nanotube. The smallest and second smallest
    transitions, $E_{11}$ and $E_{22}$, are indicated. We have used a
    broadening parameter of $\Gamma = 0.01\,$eV. }
  \label{fig9}
\end{figure}

For $2\mu$m nanotubes perpendicular to the fiber, the absorption
probabilities as defined by Eq.~(\ref{eq23e}) for different
wavelengths of light and different zigzag nanotubes are shown in
Fig.~\ref{fig9}. We do not consider the absorption of photons with
energies greater than 6eV since these are not visible and require the
addition of the higher energy $\sigma$ orbitals for accurate results.
Distinct absorption peaks are clearly visible and the largest
absorption occurs for a (11,0) nanotube.  The absorption for a
nanotube in a linearly polarized coherent plane wave [see
Eq.~(\ref{eq25b})] is also shown in Fig.~\ref{fig9}. This beam has a
cross-sectional area of $4\mu m^2$ and exhibits the same absorption
peaks as the fiber, but varies less with the light's frequency. It can
be seen that the (11,0) nanotube has its smallest energy transition
dramatically reduced. This extra effect is caused due to larger
evanescent fields, for an increasing wavelength relative to the fiber
radius. This reduces the field intensity and absorption. The quantum
efficiencies are a similar order of magnitude as those observed
experimentally for plane waves~\cite{doi:10.1021/nl034313e,PhysRevLett.93.037404,PhysRevLett.101.077402}.
The different nanotubes show shifted absorption peaks. These can be
further adjusted with external fields or choosing other
nanotubes~\cite{PhysRevLett.106.156809,PhysRevB.64.153404,0953-8984-17-37-019}.
The resonant energy values are unchanged with the orientation and this
allows us to choose a range to average over as a general measure of
absorption. We chose to calculate the mean absorption $\bar\eta$ over
the (7,0) nanotube's lowest absorption energy, particularly we chose a
range of 1eV from 1.3eV (953nm) to 2.3eV (539nm). The resulting
$\bar\eta $ is approximately independent of $\Gamma$ in a range of
$\Gamma = 0.01$eV to $\Gamma = 0.001$eV deviating only by a few
percent.

\begin{figure}[t]
  \centering\includegraphics[width=8cm]{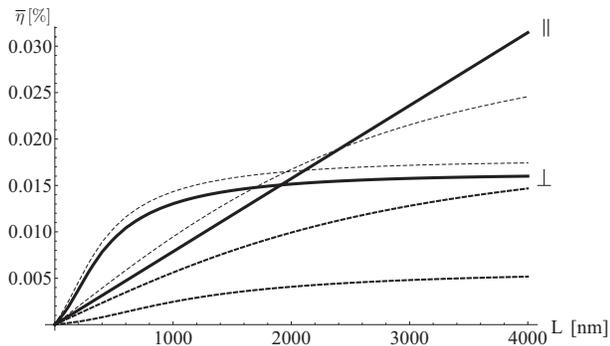}
  \caption{%
    Average photon absorption probability, $\overline{\eta}$, for a straight (7,0)
    nanotube of length $L$. At $L=4\mu$m, from top to bottom, the angles between nanotube and
    fiber are $\phi=0$ (parallel),$\,-\pi/32,\,-3\pi/8,\,\pi/2$ (perpendicular),$\,\pi/32,\,\pi/8$. The mean absorption has taken over a $1$eV region, from $1.3$eV to $2.3$eV.}
  \label{fig8}
\end{figure}

The corresponding mean absorption against nanotube length, for the
lowest energy transition, is shown in Fig.~\ref{fig8} for various
angles between the straight nanotube and nanofiber. The results show
that the absorption converges to a maximum value as the length is
increased, unless the nanotube is parallel to the fiber. The nanotube
perpendicular to the fiber has a very strong absorption for short
lengths. In this situation we see the absorption increasing strongly
with nanotube length which is due to the linear increase in electron
number. As the length increases further this effect is counterbalanced
by the fact that the field strength decreases exponentially away from
the nanofiber. The absorption hardly increases at all after the
nanotube exceeds approximately $2\,\mu$m. However, over these short distances the absorption of the perpendicular nanotube can be improved upon by shifting the nanotube slightly away from a perfectly perpendicular arrangement. The parallel orientation increases
slowly but does not peak. This effect will be discussed later in this
section and we will find that the probability can be enhanced by
spiralling the nanotube to combine both effects. If linear polarized
light was used instead of circular polarized light the absorption
could be twice as high depending on the nanotube's position in the
nanofiber plane. We also see a difference between angles of $\pm
\pi/32$, with the higher absorption being dependent on the light's
polarization and propagation.

\begin{figure}[t]
  \centering\includegraphics[width=8cm]{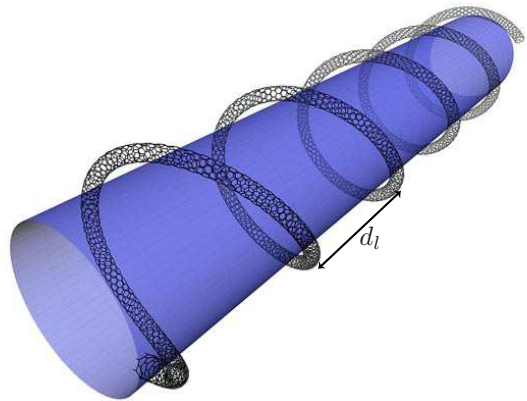}
  \caption{%
  (Color online) Nanotube spiralling around a nanofiber. The length of one loop, along the z axis, is labeled as $d_l$.}
  \label{fig10}
\end{figure}

The strong absorption for a perpendicular nanotube is limited by the
drop-off in field strength. However, this can be prevented by
maintaining a constant distance between the nanotube and nanofiber
center, $R_n$. The nanotube can locally approximate a perpendicular
nanotube by spiralling around the nanofiber, as illustrated in
Fig.~\ref{fig10}. Although this bending does alter the electronic and
optical properties these effects are small and can be safely ignored
here~\cite{PhysRevB.82.193409}. We define a `winding number', $W$, for
the spiral as the number of loops per unit length along the z axis.
This winding number is equal to $W=1/d_l$ where $d_l$ is the
z-distance for one loop. An angle is also formed between the
spiralling nanotube and the nanofiber's direction, which is given by
$\Phi_s=\arctan(2\pi W R_n)$. Since these spiralling nanotubes can
interact with the field over an arbitrary length their absorption's
approach $100\%$ given sufficient length and an allowed transition.

%
\begin{figure}[t]
  \centering\includegraphics[width=8cm]{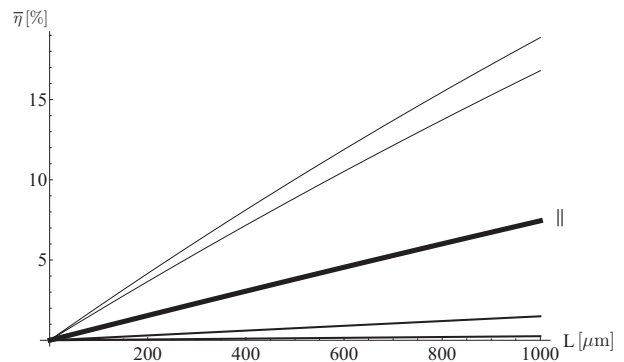}
  \caption{%
    Average photon absorption probability, $\overline{\eta}$, for a
    (7,0) nanotube of length $L$ coiled around the fiber. The average
    is taken from $1.3$eV to $2.3$eV. From top to bottom the winding
    numbers are $-0.0016\,\mathrm{nm}^{-1}$,
    $-0.0008\,\mathrm{nm}^{-1}$, $0\,\mathrm{nm}^{-1}$(Parallel),
    $0.0016\,\mathrm{nm}^{-1}$ and $0.0008\,\mathrm{nm}^{-1}$.}
  \label{fig11}
\end{figure}

\begin{figure}[t]
  \centering\includegraphics[width=8cm]{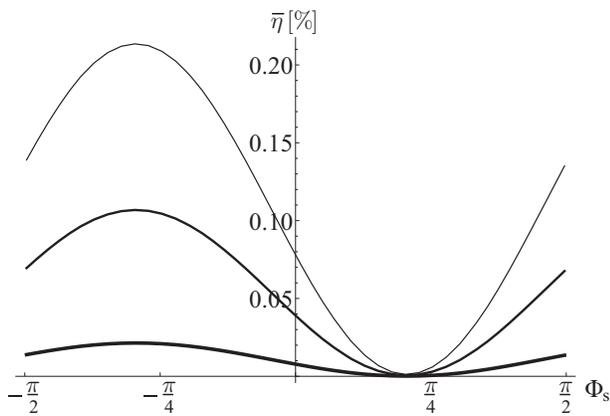}
  \caption{%
Average photon absorption probability, $\overline{\eta}$, for (7,0) nanotubes coiled around the fiber for different winding parameters. From top to bottom the nanotubes have lengths $10000$nm, $5000$nm and $1000$nm. The average is taken from $1.3$eV to $2.3$eV.}
  \label{fig12}
\end{figure}

The average absorption probabilities in this case are plotted in
Fig.~\ref{fig11} and show a steady increase in the absorption
probability with nanotube length. The parallel nanotube is also shown.
This demonstrates that the spiralling nanotubes can have higher
absorption probabilities than the parallel configuration. In
Fig.~\ref{fig12} we have plotted the average absorption probability
against $\Phi_s$ for nanotubes of different lengths. An optimal
spiralling rate to enhance the absorption can be seen. We found that the optimal value of this winding rate is $W_{opt} = e^m_{\varphi}/(2\pi R_n e^m_z)$. This was obtained by maximizing the alignment between the nanotube and $\mathbf{e}^{m}$.

\section{Applications \label{temp-deviceapp}}

The nanotube-nanofiber setups discussed in the previous sections open
up possibilities for a range of applications, particularly highly
sensitive photodetectors. These systems would detect light guided
within a fiber. In this section we discuss the possibilities. Note that we have only considered the quantum efficiency of the absorption and that detection of the charge excitations is beyond the scope of this paper. However, certain nanotube properties such as ballistic electron transport and low capacitance should be a great advantage for this detection. The bandgap of carbon nanotubes decreases with the
nanotube size, so for optical and near infrared wavelengths small
diameter nanotubes are required. This rules out the possibility of
encasing a nanofiber within a nanotube. Instead, a practical setup is
given by arranging $N$ horizontal nanotubes in a parallel array and
placing the nanofiber orthogonally on top of the array. Based on
current nanotube arrays, the density of nanotubes would be $~1-100$
nanotubes per $\mu m$ ~\cite{doi:10.1021/nl803496s,springerlink:10.1007/s12274-010-0054-0,doi:10.1021/nn201314t,ADMA:ADMA200903238,doi:10.1021/nn102305z}. We will use $\overline{\eta}$ as a measure of
the absorption probability for one nanotube. The photon absorption
probability of each nanotube is then, in case of a (7,0) nanotube,
given by the top line in Fig.~\ref{fig8} and the overall absorption
probability is
\begin{equation}
\eta_{tot} = 1-(1-\overline{\eta})^N.
\label{eq:prob}
\end{equation}
Taking $\overline{\eta}=0.00015$ (see Fig.~\ref{fig8}) this leads to
$\eta_{tot}>95$\% for $N>20000$, a value greatly exceeding those of
currently available APDs~\cite{Hadfield}. For $N>40000$ the efficiency
exceeds $99$\% which can currently only be achieved by highly complex
superconducting detectors~\cite{Hadfield}. The advantage of our setup
is that it can be operated at room temperature. Each nanotube would
require a length of $~2\mu$m and has to be connected at the ends by
electrodes~\cite{0957-4484-22-24-245305} which collect the excited electrons via an applied voltage.
Although this should be possible in the near future, current
technology cannot generate an array of unique nanotubes.

\begin{figure}[t]
  \centering\includegraphics[width=8cm]{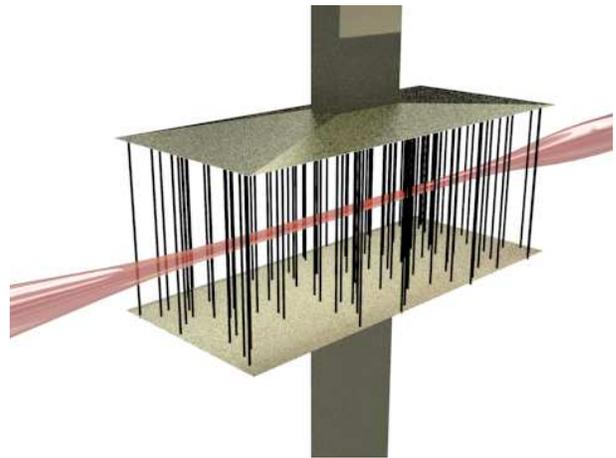}
  \caption{%
    (Color online) Illustration of a possible photodetector. Here a
    `forest' of aligned semiconducting nanotubes (thick dark lines)
    are grown between two electrodes. The nanofiber is positioned
    between these electrodes. Once a light field excites an electron
    the resistance between the electrodes drops dramatically, which
    allows the photon to be recorded.}
  \label{fig12-temp}
\end{figure}

\begin{figure}[]
  \centering\includegraphics[width=8cm]{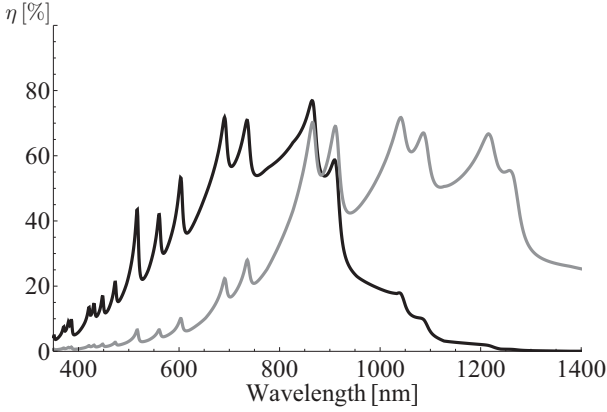}
  \caption{%
Absorption probability for a circularly polarized photon in a nanofiber laid inside a `forest' of nanotubes (see Fig.~\ref{fig12-temp}). This is given for  $2\,\mu$m long vertically aligned nanotubes. The nanotubes are in a region that extends $500$nm away from the fiber and $15\mu m$ along its length. The density of the array is taken as 900 nanotubes per $\mu\mathrm{m}^2$. The fibers diameter is taken to be either $250\,$nm (black line) and $400\,$nm (gray line). A broadening parameter of $\Gamma = 0.01\,$eV was used.
    }
  \label{fig13-temp}
\end{figure}

Aligned vertical nanotubes can also be grown on a conducting
substrate, that can then serve as one electrode. The nanofiber can
then be placed orthogonally to the nanotubes and the remaining ends of
the nanotubes connected to an additional electrode (see
Fig.~\ref{fig12-temp}).  The diameter of the nanotubes in this case
can be in the range of $1\pm0.5\,$nm~\cite{doi:10.1021/ja060944+,doi:10.1021/jp8073877}. Recently,
progress has been made in the generation of such semiconducting
nanotube `forests', although a semiconducting nanotube purity of
$100\%$ has yet to be achieved
reliably~\cite{C1JM10399G,doi:10.1021/nl800967n,Collins27042001}. These nanotube
systems typically have a density of $~10-10000$ nanotubes per $\mu
\mathrm{m}^{2}$ ~\cite{Zhong20062009,Murakami2004298,doi:10.1021/nn1025675,Ibrahim20115029}. The nanotubes are distributed uniformly over a selected region
and we assume that they are a uniform mix of semiconducting zigzag
nanotubes with a diameter in the range $1\pm0.5\,$nm. We calculated the
overall absorption probability when we have a forest that extends a
distance of 500nm from the nanofiber and $15\,\mu$m along its length,
with a density of $900$ nanotubes per $\mu \mathrm{m}^2$. The results with
nanofibers that have diameters of $250\,$nm and $400\,$nm are shown in
Fig.~\ref{fig13-temp}. These absorb light of a wide range of
wavelengths, that can be selected by the nanotubes present and choice
of nanofiber diameter. A typical absorption probability of
$\eta_{tot}>50\%$ can be seen, for $250\,$nm diameter fibers, and by
extending the system's length from $15\,\mu$m to $50\,\mu$m this is
increased to $\eta_{tot}>95\%$. Nanotubes around a $400\,$nm fiber are
also seen to absorb light at wavelengths that are typically used for
optical communication. Due to the nanotube's bandgap dependence on
external fields there is also the possibility of adjusting the
absorption frequencies by using an external field.

An additional possible setup is given by arranging the nanofiber and
the nanotube parallel to each other. Taking 100 nanotubes of length
$L=1\,$mm parallel to the fiber and using
$\overline{\eta}=0.07$ (see Fig.~\ref{fig11}) we obtain an overall absorption probability of
$\eta_{tot}>99\%$ which again greatly exceeds that of standard APDs.


As a final setup we consider the coil geometry shown in
Fig.~\ref{fig10} which has a high absorption probability of up to
100\% for long nanotubes. However, producing such a setup in a
laboratory is rather challenging with current technology. This setup
also allows for further specification of the absorbed light's
polarization or propagation direction with the choice of winding
number. The winding also dramatically reduces the length of the
system. For a nanotube with a winding number of $W=-0.1\,\mathrm{nm}^{-1}$, the average
absorption between 1.5eV and 2.5eV exceeds $50\%$ when the nanotube's
length is $5\,$mm. For the $250\,$nm diameter fiber this only extends
$64\,\mu$m along the fiber.

\section{Summary \label{sec5}}
In this paper we have calculated the probability of
absorbing a photon with zigzag carbon nanotubes. The light field is
allowed to vary along the nanotube, i.e. no dipole approximation is made, which has enabled us to treat the absorption
of light from optical nanofibers. We found that there is a strong
dependence on the system's geometry and have devised setups for high absorption. If we spiral the nanotube around
the fiber, we find that an absorption of circularly polarized light, arbitrarily close to 100\% can be achieved. For straight nanotubes, that are not parallel to the fiber, we find the absorption probability converges as the nanotube's length increases. We have found a simple expression for the absorption probability, which is independent of the coherence length. Currently, the coherence lengths of carbon nanotubes is
an area of extensive research with results ranging from $10\,$nm, at
room temperature, to a few microns~\cite{PhysRevB.80.081410,PhysRevLett.95.076803,ISI:A1997WR25600045,PhysRevB.72.113410,PhysRevLett.94.086802}. They seem to be highly dependent on the temperature, impurities, defects
and surrounding fields. Once excited, the radiative lifetimes of the
excitations have been observed to range from $3$ to
$100\,$ns~\cite{PhysRevB.80.081410,PhysRevLett.92.177401}. The nonradiative decay seems
to be much faster, of the order of a few picoseconds~\cite{PhysRevLett.90.057404}. Introducing excitons into the model and analyzing the dynamics of quantized single photon states within a carbon nanotube is an
interesting area of further research. Yet, the results here are an
important step towards calculating the absorptions within nanostructures and are of importance to future nanotube optoelectonics.

The research of UD was supported by the Centre for Quantum Technologies, National University of Singapore and the ESF via EuroQUAM (EPSRC Grant No. EP/E041612/1). SB acknowledges support from the \mbox{EPSRC} Doctoral Training Accounts.


\appendix

\section{Calculating the Optical Matrix Element \label{app1}}

The matrix element for the interaction between can be found by
substituting in the wavefunctions to give
\begin{align} G
  &= \bra{\Psi^c (\mathbf{k}')}\mathbf{A}^+.\nabla \ket{\Psi^v (\mathbf{k})}   \notag \\
  &= \sum_{s,t=A,B} c_s^{c*}(\mathbf{k}') c_t^{v}(\mathbf{k}) \bra{\tilde{p}_z^s (\mathbf{k}')} \mathbf{A}^+.\nabla \ket{\tilde{p}_z^t (\mathbf{k})}\notag \\
  &= \frac{1}{N_{cells}} \sum_{s,t=A,B} c_s^{c*}(\mathbf{k}') c_t^{v}(\mathbf{k})  \notag\\
  & \sum_{\mathbf{r}_1\in \mathbf{R}_s,\mathbf{r}_2\in \mathbf{R}_t} \e^{\ii\mathbf{k}.\mathbf{r}_2-\ii\mathbf{k}'.\mathbf{r}_1} \bra{p_z(\mathbf{r}-\mathbf{r}_1)}  \mathbf{A}^+.\nabla \ket{p_z(\mathbf{r}-\mathbf{r}_2)}   \notag\\
  &= \frac{1}{N_G N_L} \sum_{\mathbf{r}_1\in \mathbf{R}_A,\mathbf{r}_2\in \mathbf{R}_B}  \notag\\
  & c_A^{c*}(\mathbf{k}') c_B^{v}(\mathbf{k}) \e^{\ii\mathbf{k}.\mathbf{r}_2-\ii\mathbf{k}'.\mathbf{r}_1} \notag \\
  & \qquad\qquad\qquad \bra{p_z(\mathbf{r}-\mathbf{r}_1)} \mathbf{A}^+.\nabla \ket{p_z(\mathbf{r}-\mathbf{r}_2)}   \notag\\
  & + c_B^{c*}(\mathbf{k}') c_A^{v}(\mathbf{k}) \e^{\ii\mathbf{k}.\mathbf{r}_1-\ii\mathbf{k}'.\mathbf{r}_2} \notag \\
  & \qquad\qquad\qquad \bra{p_z(\mathbf{r}-\mathbf{r}_2)} \mathbf{A}^+.\nabla \ket{p_z(\mathbf{r}-\mathbf{r}_1)}  \notag\\
  &= \frac{M\sqrt{3}}{a N_G N_L} (c_A^{c*}(\mathbf{k}') c_B^{v}(\mathbf{k}) \notag \\
  & \qquad\qquad\qquad \sum_{\mathbf{r}_1\in \mathbf{R}_A} \e^{-\ii(\mathbf{k}'-\mathbf{k}).\mathbf{r}_1} \mathbf{A}^+(\mathbf{r}_1).\mathbf{v}^A (\mathbf{k})  \notag \\
  &+ c_B^{c*}(\mathbf{k}') c_A^{v}(\mathbf{k}) \notag \\
  & \qquad\qquad\qquad \sum_{\mathbf{r}_2\in R_B} \e^{-\ii(\mathbf{k}'-\mathbf{k}).\mathbf{r}_2}\mathbf{A}^+(\mathbf{r}_2).\mathbf{v}^B (\mathbf{k})).
  \label{eq27}
\end{align}

Here we have assumed that the orbitals are symmetric and that $\mathbf{A}^+$ is constant across each of the nanotube's unit cells. The expression for $G$ can then be split into separate unit cells and
directions
\begin{align}
  G &= G_x + G_y + G_z   \\
  G_z &= \frac{M\sqrt{3}}{2 a n N_L} ( \sum_{l=1}^{N_L} \e^{\ii (al\sqrt{3}-(L/2)) (k_{||}-k'_{||})} \notag \\
  & \qquad\qquad\qquad\qquad\qquad\qquad A^+_z(la\sqrt{3}-(L/2)) )  \notag \\
  & \sum_{j=1}^{n} (c_A^{c*}(\mathbf{k}') c_B^{v}(\mathbf{k}) \e^{-\ii j a (k'_{\bot}-k_{\bot})}  \notag \\
  & (1+\e^{-\ii a (\mathbf{k}'-\mathbf{k}).(\sqrt{3}/2,1/2)}) v_z^A (\mathbf{k})   \notag \\
  & - c_B^{c*}(\mathbf{k}') c_A^{v}(\mathbf{k}) \e^{-\ii j a (k'_{\bot}-k_{\bot})} \e^{\ii a (k_{||}-k'_{||})/\sqrt{3}}  \notag \\
  & (1+\e^{-\ii a (\mathbf{k}'-\mathbf{k}).(\sqrt{3}/2,1/2)}) v_z^A (\mathbf{k})^*) \notag \\
  &= \frac{1}{N_L} D_z ( \sum_{l=1}^{N_L} \e^{\ii(k_{||}-k'_{||})(la\sqrt{3}-(L/2))} \notag \\
  & \qquad\qquad\qquad\qquad\qquad\qquad A^+_z(la\sqrt{3}-(L/2)) ).
  \label{eq29}
\end{align}

In order to calculate $G_x$ and $G_y$ we must take the curvature of
the nanotube into consideration. To do this we use the method from  Ref.~\cite{Jiang20043169} and introduce the
parameters
\begin{align}
  v^{A_0}_{\pm} &= \e^{\ii a \mathbf{k}.(-1/(2\sqrt{3}),-1/2)} (\e^{\mp 2\pi \ii /2n}-1)  \notag \\
  &+ \e^{\ii a \mathbf{k}.{-1/(2\sqrt{3},1/2)}} (\e^{\pm 2\pi \ii /2n}-1) \\
  v^{B_0}_{\pm} &= \e^{\ii a \mathbf{k}.(1/(2\sqrt{3}),-1/2)} (\e^{\mp 2\pi \ii /2n}-1)  \notag \\
  &+ \e^{\ii a \mathbf{k}.(1/(2\sqrt{3}),1/2)} (\e^{\pm 2\pi \ii /2n}-1) \\
  v^{A}_{x(\theta)} (\mathbf{k}) &= R_t \frac{\e^{\ii\theta} v^{A_0}_{+} + \e^{-\ii\theta} v^{A_0}_{-}}{2} \\
  v^{B}_{x(\theta)} (\mathbf{k}) &= R_t \frac{\e^{\ii\theta} v^{B_0}_{+} + \e^{-\ii\theta} v^{B_0}_{-}}{2} \\
  v^{A}_{y(\theta)} (\mathbf{k}) &= R_t \frac{\e^{\ii\theta} v^{A_0}_{+} + \e^{-\ii\theta} v^{A_0}_{-}}{2\ii}\\
  v^{B}_{y(\theta)} (\mathbf{k}) &= R_t \frac{\e^{\ii\theta} v^{B_0}_{+} + \e^{-\ii\theta} v^{B_0}_{-}}{2\ii}.
  \label{eq30}
\end{align}
From these we calculate $G_{d=x,y}$ to be

\begin{align}
  G_d &= \frac{M\sqrt{3}}{2 a n N_L} ( \sum_{l=1}^{N_L} \e^{\ii(k_{||}-k'_{||})(la\sqrt{3}-(L/2))} \notag \\
  & \qquad\qquad\qquad\qquad\qquad\qquad A^+_d(la\sqrt{3}-(L/2)) )  \notag \\
  & \sum_{j=1}^{n} c_A^{c*}(\mathbf{k}') c_B^{v}(\mathbf{k}) (v^{A}_{d(2\pi j/n)}(\mathbf{k}) \e^{-\ii j a(k'_{\bot}-k_{\bot})}+ \notag \\
  & + v^{A}_{d(2\pi (j+1/2)/n)}(\mathbf{k}) \e^{-\ii j a(k'_{\bot}-k_{\bot})} \e^{-\ii a \mathbf{k}'.(1/(2\sqrt{3}),1/2)}) \notag \\
  & - c_B^{c*}(\mathbf{k}') c_A^{v}(\mathbf{k}) (v^{B}_{d(2\pi j/n)}(\mathbf{k}) \e^{-\ii j a(k'_{\bot}-k_{\bot})} \e^{-\ii a k'_{||}/\sqrt{3}} \notag \\
  & + v^{B}_{d(2\pi (j+1/2)/n)}(\mathbf{k}) \e^{-\ii j a(k'_{\bot}-k_{\bot})} \e^{-\ii a
    \mathbf{k}'.(5/(2\sqrt{3}),1/2)}) \notag \\
  &= \frac{1}{N_L} \mathbf{D}_d ( \sum_{l=1}^{N_L} \e^{\ii(k_{||}-k'_{||})(la\sqrt{3}-(L/2))} \notag \\
  & \qquad\qquad\qquad\qquad\qquad\qquad A^+_d(la\sqrt{3}-(L/2)) ).
  \label{eq31}
\end{align}

Hence, we obtain
\begin{equation}
  G = \frac{1}{N_L} \mathbf{D} \sum_{l=1}^{N_L} \e^{\ii(k_{||}-k'_{||})(la\sqrt{3}-(L/2))} \mathbf{A}^+(la\sqrt{3}-(L/2)).
\end{equation}

\section{Classical Nanofiber Field Modes \label{app2}}
For light of wavelength, $\lambda$, and $k=2\pi/\lambda$, the field parameters must satisfy the fiber eigenvalue equation~\cite{snyder1983optical}
\begin{align}
  \frac{J_0(hR)}{hR J_1 (hR)} &= -\frac{n_1^2+n_2^2}{2n_1^2} \frac{K_1'(qR)}{qR K_1(qR)}+\frac{1}{h^2 R^2}  \notag\\
  & -[ \left( \frac{n_1^2-n_2^2}{2n_1^2} \frac{K_1'(qR)}{qR K_1(qR)} \right)^2  \notag\\
  & +\frac{\beta^2}{n_1^2 k^2} \left( \frac{1}{q^2 R^2} +\frac{1}{h^2 R^2} \right)^2 ]^{1/2},
  \label{eq32}
\end{align}
with $J_{\nu}$ referring to the Bessel function of the first kind and $K_{\nu}$ being the modified Bessel function of the second kind. By numerically solving Eq.~(\ref{eq32}) the
value of the propagation constant, $\beta$, is determined. We also define the parameters
$h=(n_1^2 k^2 - \beta^2)^{1/2}$, $q=(\beta^2- n_2^2 k^2)^{1/2}$ and
\begin{equation}
  g=\left( \frac{1}{q^2 R^2} +\frac{1}{h^2 R^2}  \right) / \left( \frac{J_1'(hR)}{hRJ_1(hR)} +\frac{K_1'(qR)}{qRK_1(qR)} \right).
\end{equation}

Inside the fiber ($0<r<R$) the guided mode, $m=(f,p)$, have the form
\begin{align}
  e^m_r &= \ii \frac{q K_1 (R)}{h J_1 (hR)}[(1-g)J_0 (hr) - (1+g)J_2(hr)], \\
  e^m_{\varphi} &= -p\frac{q K_1 (qR)}{h J_1 (hR)}[(1-g)J_0 (hr) + (1+g)J_2(hr)], \\
  e^m_z &= f\frac{2q K_1 (qR)}{\beta J_1 (hR)}J_1 (hr),
\end{align}
and outside $r > R$
\begin{align}
  e^m_r &= \ii [(1-g)K_0 (qr) + (1+g)K_2(qr)], \\
  e^m_{\varphi} &= -p[(1-g)K_0 (qr) - (1+g)K_2(qr)], \\
  e^m_z &= f\frac{2q }{\beta }K_1 (qr),
\end{align}

These are normalized with a factor given by
\begin{equation}
  \int_0^{2\pi} \int_0^{R} n_1^2 \left|\mathbf{e}\right|^2 r dr d\varphi + \int_0^{2\pi} \int_R^{\infty} n_2^2 \left|\mathbf{e}\right|^2 r dr d\varphi = A.
\end{equation}


\begin{thebibliography}{76}%
\makeatletter
\providecommand \@ifxundefined [1]{%
 \@ifx{#1\undefined}
}%
\providecommand \@ifnum [1]{%
 \ifnum #1\expandafter \@firstoftwo
 \else \expandafter \@secondoftwo
 \fi
}%
\providecommand \@ifx [1]{%
 \ifx #1\expandafter \@firstoftwo
 \else \expandafter \@secondoftwo
 \fi
}%
\providecommand \natexlab [1]{#1}%
\providecommand \enquote  [1]{``#1''}%
\providecommand \bibnamefont  [1]{#1}%
\providecommand \bibfnamefont [1]{#1}%
\providecommand \citenamefont [1]{#1}%
\providecommand \href@noop [0]{\@secondoftwo}%
\providecommand \href [0]{\begingroup \@sanitize@url \@href}%
\providecommand \@href[1]{\@@startlink{#1}\@@href}%
\providecommand \@@href[1]{\endgroup#1\@@endlink}%
\providecommand \@sanitize@url [0]{\catcode `\\12\catcode `\$12\catcode
  `\&12\catcode `\#12\catcode `\^12\catcode `\_12\catcode `\%12\relax}%
\providecommand \@@startlink[1]{}%
\providecommand \@@endlink[0]{}%
\providecommand \url  [0]{\begingroup\@sanitize@url \@url }%
\providecommand \@url [1]{\endgroup\@href {#1}{\urlprefix }}%
\providecommand \urlprefix  [0]{URL }%
\providecommand \Eprint [0]{\href }%
\providecommand \doibase [0]{http://dx.doi.org/}%
\providecommand \selectlanguage [0]{\@gobble}%
\providecommand \bibinfo  [0]{\@secondoftwo}%
\providecommand \bibfield  [0]{\@secondoftwo}%
\providecommand \translation [1]{[#1]}%
\providecommand \BibitemOpen [0]{}%
\providecommand \bibitemStop [0]{}%
\providecommand \bibitemNoStop [0]{.\EOS\space}%
\providecommand \EOS [0]{\spacefactor3000\relax}%
\providecommand \BibitemShut  [1]{\csname bibitem#1\endcsname}%
\let\auto@bib@innerbib\@empty
\bibitem [{\citenamefont {Charlier}\ \emph {et~al.}(2007)\citenamefont
  {Charlier}, \citenamefont {Blase},\ and\ \citenamefont
  {Roche}}]{RevModPhys.79.677}%
  \BibitemOpen
  \bibfield  {author} {\bibinfo {author} {\bibfnamefont {J.-C.}\ \bibnamefont
  {Charlier}}, \bibinfo {author} {\bibfnamefont {X.}~\bibnamefont {Blase}}, \
  and\ \bibinfo {author} {\bibfnamefont {S.}~\bibnamefont {Roche}},\ }\href
  {\doibase 10.1103/RevModPhys.79.677} {\bibfield  {journal} {\bibinfo
  {journal} {Rev. Mod. Phys.}\ }\textbf {\bibinfo {volume} {79}},\ \bibinfo
  {pages} {677} (\bibinfo {year} {2007})}\BibitemShut {NoStop}%
\bibitem [{\citenamefont {Ando}(2005)}]{JPSJ.74.777}%
  \BibitemOpen
  \bibfield  {author} {\bibinfo {author} {\bibfnamefont {T.}~\bibnamefont
  {Ando}},\ }\href {\doibase 10.1143/JPSJ.74.777} {\bibfield  {journal}
  {\bibinfo  {journal} {Journal of the Physical Society of Japan}\ }\textbf
  {\bibinfo {volume} {74}},\ \bibinfo {pages} {777} (\bibinfo {year}
  {2005})}\BibitemShut {NoStop}%
\bibitem [{\citenamefont {Saito}\ \emph {et~al.}(1998)\citenamefont {Saito},
  \citenamefont {Dresselhaus},\ and\ \citenamefont
  {Dresselhaus}}]{saito1998physical}%
  \BibitemOpen
  \bibfield  {author} {\bibinfo {author} {\bibfnamefont {R.}~\bibnamefont
  {Saito}}, \bibinfo {author} {\bibfnamefont {G.}~\bibnamefont {Dresselhaus}},
  \ and\ \bibinfo {author} {\bibfnamefont {M.}~\bibnamefont {Dresselhaus}},\
  }\href@noop {} {\emph {\bibinfo {title} {Physical properties of carbon
  nanotubes}}}\ (\bibinfo  {publisher} {Imperial College Press},\ \bibinfo
  {year} {1998})\BibitemShut {NoStop}%
\bibitem [{\citenamefont {Wang}\ \emph {et~al.}(2009)\citenamefont {Wang},
  \citenamefont {Li}, \citenamefont {Xie}, \citenamefont {Jin}, \citenamefont
  {Wang}, \citenamefont {Li}, \citenamefont {Jiang},\ and\ \citenamefont
  {Fan}}]{doi:10.1021/nl901260b}%
  \BibitemOpen
  \bibfield  {author} {\bibinfo {author} {\bibfnamefont {X.}~\bibnamefont
  {Wang}}, \bibinfo {author} {\bibfnamefont {Q.}~\bibnamefont {Li}}, \bibinfo
  {author} {\bibfnamefont {J.}~\bibnamefont {Xie}}, \bibinfo {author}
  {\bibfnamefont {Z.}~\bibnamefont {Jin}}, \bibinfo {author} {\bibfnamefont
  {J.}~\bibnamefont {Wang}}, \bibinfo {author} {\bibfnamefont {Y.}~\bibnamefont
  {Li}}, \bibinfo {author} {\bibfnamefont {K.}~\bibnamefont {Jiang}}, \ and\
  \bibinfo {author} {\bibfnamefont {S.}~\bibnamefont {Fan}},\ }\href {\doibase
  10.1021/nl901260b} {\bibfield  {journal} {\bibinfo  {journal} {Nano Letters}\
  }\textbf {\bibinfo {volume} {9}},\ \bibinfo {pages} {3137} (\bibinfo {year}
  {2009})}\BibitemShut {NoStop}%
\bibitem [{\citenamefont {White}\ and\ \citenamefont
  {Todorov}(1998)}]{ISI:000073761000044}%
  \BibitemOpen
  \bibfield  {author} {\bibinfo {author} {\bibfnamefont {C.}~\bibnamefont
  {White}}\ and\ \bibinfo {author} {\bibfnamefont {T.}~\bibnamefont
  {Todorov}},\ }\href {\doibase 10.1038/30420} {\bibfield  {journal} {\bibinfo
  {journal} {Nature}\ }\textbf {\bibinfo {volume} {393}},\ \bibinfo {pages}
  {240} (\bibinfo {year} {1998})}\BibitemShut {NoStop}%
\bibitem [{\citenamefont {Klinovaja}\ \emph {et~al.}(2011)\citenamefont
  {Klinovaja}, \citenamefont {Schmidt}, \citenamefont {Braunecker},\ and\
  \citenamefont {Loss}}]{PhysRevLett.106.156809}%
  \BibitemOpen
  \bibfield  {author} {\bibinfo {author} {\bibfnamefont {J.}~\bibnamefont
  {Klinovaja}}, \bibinfo {author} {\bibfnamefont {M.~J.}\ \bibnamefont
  {Schmidt}}, \bibinfo {author} {\bibfnamefont {B.}~\bibnamefont {Braunecker}},
  \ and\ \bibinfo {author} {\bibfnamefont {D.}~\bibnamefont {Loss}},\ }\href
  {\doibase 10.1103/PhysRevLett.106.156809} {\bibfield  {journal} {\bibinfo
  {journal} {Phys. Rev. Lett.}\ }\textbf {\bibinfo {volume} {106}},\ \bibinfo
  {pages} {156809} (\bibinfo {year} {2011})}\BibitemShut {NoStop}%
\bibitem [{\citenamefont {Kim}\ and\ \citenamefont
  {Chang}(2001)}]{PhysRevB.64.153404}%
  \BibitemOpen
  \bibfield  {author} {\bibinfo {author} {\bibfnamefont {Y.-H.}\ \bibnamefont
  {Kim}}\ and\ \bibinfo {author} {\bibfnamefont {K.~J.}\ \bibnamefont
  {Chang}},\ }\href {\doibase 10.1103/PhysRevB.64.153404} {\bibfield  {journal}
  {\bibinfo  {journal} {Phys. Rev. B}\ }\textbf {\bibinfo {volume} {64}},\
  \bibinfo {pages} {153404} (\bibinfo {year} {2001})}\BibitemShut {NoStop}%
\bibitem [{\citenamefont {Pacheco}\ \emph {et~al.}(2005)\citenamefont
  {Pacheco}, \citenamefont {Barticevic}, \citenamefont {Rocha},\ and\
  \citenamefont {Latg\'{e}}}]{0953-8984-17-37-019}%
  \BibitemOpen
  \bibfield  {author} {\bibinfo {author} {\bibfnamefont {M.}~\bibnamefont
  {Pacheco}}, \bibinfo {author} {\bibfnamefont {Z.}~\bibnamefont {Barticevic}},
  \bibinfo {author} {\bibfnamefont {C.~G.}\ \bibnamefont {Rocha}}, \ and\
  \bibinfo {author} {\bibfnamefont {A.}~\bibnamefont {Latg\'{e}}},\ }\href
  {\doibase 10.1088/0953-8984/17/37/019} {\bibfield  {journal} {\bibinfo
  {journal} {Journal of Physics: Condensed Matter}\ }\textbf {\bibinfo {volume}
  {17}},\ \bibinfo {pages} {5839} (\bibinfo {year} {2005})}\BibitemShut
  {NoStop}%
\bibitem [{\citenamefont {Takesue}\ \emph {et~al.}(2006)\citenamefont
  {Takesue}, \citenamefont {Haruyama}, \citenamefont {Kobayashi}, \citenamefont
  {Chiashi}, \citenamefont {Maruyama}, \citenamefont {Sugai},\ and\
  \citenamefont {Shinohara}}]{PhysRevLett.96.057001}%
  \BibitemOpen
  \bibfield  {author} {\bibinfo {author} {\bibfnamefont {I.}~\bibnamefont
  {Takesue}}, \bibinfo {author} {\bibfnamefont {J.}~\bibnamefont {Haruyama}},
  \bibinfo {author} {\bibfnamefont {N.}~\bibnamefont {Kobayashi}}, \bibinfo
  {author} {\bibfnamefont {S.}~\bibnamefont {Chiashi}}, \bibinfo {author}
  {\bibfnamefont {S.}~\bibnamefont {Maruyama}}, \bibinfo {author}
  {\bibfnamefont {T.}~\bibnamefont {Sugai}}, \ and\ \bibinfo {author}
  {\bibfnamefont {H.}~\bibnamefont {Shinohara}},\ }\href {\doibase
  10.1103/PhysRevLett.96.057001} {\bibfield  {journal} {\bibinfo  {journal}
  {Phys. Rev. Lett.}\ }\textbf {\bibinfo {volume} {96}},\ \bibinfo {pages}
  {057001} (\bibinfo {year} {2006})}\BibitemShut {NoStop}%
\bibitem [{\citenamefont {Noffsinger}\ and\ \citenamefont
  {Cohen}(2011)}]{PhysRevB.83.165420}%
  \BibitemOpen
  \bibfield  {author} {\bibinfo {author} {\bibfnamefont {J.}~\bibnamefont
  {Noffsinger}}\ and\ \bibinfo {author} {\bibfnamefont {M.~L.}\ \bibnamefont
  {Cohen}},\ }\href {\doibase 10.1103/PhysRevB.83.165420} {\bibfield  {journal}
  {\bibinfo  {journal} {Phys. Rev. B}\ }\textbf {\bibinfo {volume} {83}},\
  \bibinfo {pages} {165420} (\bibinfo {year} {2011})}\BibitemShut {NoStop}%
\bibitem [{\citenamefont {Kasumov}\ \emph {et~al.}(1999)\citenamefont
  {Kasumov}, \citenamefont {Deblock}, \citenamefont {Kociak}, \citenamefont
  {Reulet}, \citenamefont {Bouchiat}, \citenamefont {Khodos}, \citenamefont
  {Gorbatov}, \citenamefont {Volkov}, \citenamefont {Journet},\ and\
  \citenamefont {Burghard}}]{Kasumov28051999}%
  \BibitemOpen
  \bibfield  {author} {\bibinfo {author} {\bibfnamefont {A.~Y.}\ \bibnamefont
  {Kasumov}}, \bibinfo {author} {\bibfnamefont {R.}~\bibnamefont {Deblock}},
  \bibinfo {author} {\bibfnamefont {M.}~\bibnamefont {Kociak}}, \bibinfo
  {author} {\bibfnamefont {B.}~\bibnamefont {Reulet}}, \bibinfo {author}
  {\bibfnamefont {H.}~\bibnamefont {Bouchiat}}, \bibinfo {author}
  {\bibfnamefont {I.~I.}\ \bibnamefont {Khodos}}, \bibinfo {author}
  {\bibfnamefont {Y.~B.}\ \bibnamefont {Gorbatov}}, \bibinfo {author}
  {\bibfnamefont {V.~T.}\ \bibnamefont {Volkov}}, \bibinfo {author}
  {\bibfnamefont {C.}~\bibnamefont {Journet}}, \ and\ \bibinfo {author}
  {\bibfnamefont {M.}~\bibnamefont {Burghard}},\ }\href {\doibase
  10.1126/science.284.5419.1508} {\bibfield  {journal} {\bibinfo  {journal}
  {Science}\ }\textbf {\bibinfo {volume} {284}},\ \bibinfo {pages} {1508}
  (\bibinfo {year} {1999})}\BibitemShut {NoStop}%
\bibitem [{\citenamefont {Morpurgo}\ \emph {et~al.}(1999)\citenamefont
  {Morpurgo}, \citenamefont {Kong}, \citenamefont {Marcus},\ and\ \citenamefont
  {Dai}}]{Morpurgo08101999}%
  \BibitemOpen
  \bibfield  {author} {\bibinfo {author} {\bibfnamefont {A.~F.}\ \bibnamefont
  {Morpurgo}}, \bibinfo {author} {\bibfnamefont {J.}~\bibnamefont {Kong}},
  \bibinfo {author} {\bibfnamefont {C.~M.}\ \bibnamefont {Marcus}}, \ and\
  \bibinfo {author} {\bibfnamefont {H.}~\bibnamefont {Dai}},\ }\href {\doibase
  10.1126/science.286.5438.263} {\bibfield  {journal} {\bibinfo  {journal}
  {Science}\ }\textbf {\bibinfo {volume} {286}},\ \bibinfo {pages} {263}
  (\bibinfo {year} {1999})}\BibitemShut {NoStop}%
\bibitem [{\citenamefont {Uchida}\ and\ \citenamefont
  {Okada}(2009)}]{PhysRevB.79.085402}%
  \BibitemOpen
  \bibfield  {author} {\bibinfo {author} {\bibfnamefont {K.}~\bibnamefont
  {Uchida}}\ and\ \bibinfo {author} {\bibfnamefont {S.}~\bibnamefont {Okada}},\
  }\href {\doibase 10.1103/PhysRevB.79.085402} {\bibfield  {journal} {\bibinfo
  {journal} {Phys. Rev. B}\ }\textbf {\bibinfo {volume} {79}},\ \bibinfo
  {pages} {085402} (\bibinfo {year} {2009})}\BibitemShut {NoStop}%
\bibitem [{\citenamefont {Snow}\ \emph {et~al.}(2005)\citenamefont {Snow},
  \citenamefont {Perkins}, \citenamefont {Houser}, \citenamefont {Badescu},\
  and\ \citenamefont {Reinecke}}]{Snow25032005}%
  \BibitemOpen
  \bibfield  {author} {\bibinfo {author} {\bibfnamefont {E.~S.}\ \bibnamefont
  {Snow}}, \bibinfo {author} {\bibfnamefont {F.~K.}\ \bibnamefont {Perkins}},
  \bibinfo {author} {\bibfnamefont {E.~J.}\ \bibnamefont {Houser}}, \bibinfo
  {author} {\bibfnamefont {S.~C.}\ \bibnamefont {Badescu}}, \ and\ \bibinfo
  {author} {\bibfnamefont {T.~L.}\ \bibnamefont {Reinecke}},\ }\href {\doibase
  10.1126/science.1109128} {\bibfield  {journal} {\bibinfo  {journal}
  {Science}\ }\textbf {\bibinfo {volume} {307}},\ \bibinfo {pages} {1942}
  (\bibinfo {year} {2005})}\BibitemShut {NoStop}%
\bibitem [{\citenamefont {Tang}\ \emph {et~al.}(2006)\citenamefont {Tang},
  \citenamefont {Bansaruntip}, \citenamefont {Nakayama}, \citenamefont
  {Yenilmez}, \citenamefont {Chang},\ and\ \citenamefont
  {Wang}}]{doi:10.1021/nl060613v}%
  \BibitemOpen
  \bibfield  {author} {\bibinfo {author} {\bibfnamefont {X.}~\bibnamefont
  {Tang}}, \bibinfo {author} {\bibfnamefont {S.}~\bibnamefont {Bansaruntip}},
  \bibinfo {author} {\bibfnamefont {N.}~\bibnamefont {Nakayama}}, \bibinfo
  {author} {\bibfnamefont {E.}~\bibnamefont {Yenilmez}}, \bibinfo {author}
  {\bibfnamefont {Y.-l.}\ \bibnamefont {Chang}}, \ and\ \bibinfo {author}
  {\bibfnamefont {Q.}~\bibnamefont {Wang}},\ }\href {\doibase
  10.1021/nl060613v} {\bibfield  {journal} {\bibinfo  {journal} {Nano Letters}\
  }\textbf {\bibinfo {volume} {6}},\ \bibinfo {pages} {1632} (\bibinfo {year}
  {2006})}\BibitemShut {NoStop}%
\bibitem [{\citenamefont {Freitag}\ \emph {et~al.}(2003)\citenamefont
  {Freitag}, \citenamefont {Martin}, \citenamefont {Misewich}, \citenamefont
  {Martel},\ and\ \citenamefont {Avouris}}]{doi:10.1021/nl034313e}%
  \BibitemOpen
  \bibfield  {author} {\bibinfo {author} {\bibfnamefont {M.}~\bibnamefont
  {Freitag}}, \bibinfo {author} {\bibfnamefont {Y.}~\bibnamefont {Martin}},
  \bibinfo {author} {\bibfnamefont {J.~A.}\ \bibnamefont {Misewich}}, \bibinfo
  {author} {\bibfnamefont {R.}~\bibnamefont {Martel}}, \ and\ \bibinfo {author}
  {\bibfnamefont {P.}~\bibnamefont {Avouris}},\ }\href {\doibase
  10.1021/nl034313e} {\bibfield  {journal} {\bibinfo  {journal} {Nano Letters}\
  }\textbf {\bibinfo {volume} {3}},\ \bibinfo {pages} {1067} (\bibinfo {year}
  {2003})}\BibitemShut {NoStop}%
\bibitem [{\citenamefont {Spataru}\ \emph {et~al.}(2004)\citenamefont
  {Spataru}, \citenamefont {Ismail-Beigi}, \citenamefont {Benedict},\ and\
  \citenamefont {Louie}}]{PhysRevLett.92.077402}%
  \BibitemOpen
  \bibfield  {author} {\bibinfo {author} {\bibfnamefont {C.~D.}\ \bibnamefont
  {Spataru}}, \bibinfo {author} {\bibfnamefont {S.}~\bibnamefont
  {Ismail-Beigi}}, \bibinfo {author} {\bibfnamefont {L.~X.}\ \bibnamefont
  {Benedict}}, \ and\ \bibinfo {author} {\bibfnamefont {S.~G.}\ \bibnamefont
  {Louie}},\ }\href {\doibase 10.1103/PhysRevLett.92.077402} {\bibfield
  {journal} {\bibinfo  {journal} {Phys. Rev. Lett.}\ }\textbf {\bibinfo
  {volume} {92}},\ \bibinfo {pages} {077402} (\bibinfo {year}
  {2004})}\BibitemShut {NoStop}%
\bibitem [{\citenamefont {Jiang}\ \emph {et~al.}(2004)\citenamefont {Jiang},
  \citenamefont {Saito}, \citenamefont {Gr\"{u}neis}, \citenamefont
  {Dresselhaus},\ and\ \citenamefont {Dresselhaus}}]{Jiang20043169}%
  \BibitemOpen
  \bibfield  {author} {\bibinfo {author} {\bibfnamefont {J.}~\bibnamefont
  {Jiang}}, \bibinfo {author} {\bibfnamefont {R.}~\bibnamefont {Saito}},
  \bibinfo {author} {\bibfnamefont {A.}~\bibnamefont {Gr\"{u}neis}}, \bibinfo
  {author} {\bibfnamefont {G.}~\bibnamefont {Dresselhaus}}, \ and\ \bibinfo
  {author} {\bibfnamefont {M.}~\bibnamefont {Dresselhaus}},\ }\href {\doibase
  10.1016/j.carbon.2004.07.028} {\bibfield  {journal} {\bibinfo  {journal}
  {Carbon}\ }\textbf {\bibinfo {volume} {42}},\ \bibinfo {pages} {3169 }
  (\bibinfo {year} {2004})}\BibitemShut {NoStop}%
\bibitem [{\citenamefont {Motavas}\ \emph {et~al.}(2010)\citenamefont
  {Motavas}, \citenamefont {Ivanov},\ and\ \citenamefont
  {Nojeh}}]{PhysRevB.82.085442}%
  \BibitemOpen
  \bibfield  {author} {\bibinfo {author} {\bibfnamefont {S.}~\bibnamefont
  {Motavas}}, \bibinfo {author} {\bibfnamefont {A.}~\bibnamefont {Ivanov}}, \
  and\ \bibinfo {author} {\bibfnamefont {A.}~\bibnamefont {Nojeh}},\ }\href
  {\doibase 10.1103/PhysRevB.82.085442} {\bibfield  {journal} {\bibinfo
  {journal} {Phys. Rev. B}\ }\textbf {\bibinfo {volume} {82}},\ \bibinfo
  {pages} {085442} (\bibinfo {year} {2010})}\BibitemShut {NoStop}%
\bibitem [{\citenamefont {Zarifi}\ and\ \citenamefont
  {Pedersen}(2009)}]{PhysRevB.80.195422}%
  \BibitemOpen
  \bibfield  {author} {\bibinfo {author} {\bibfnamefont {A.}~\bibnamefont
  {Zarifi}}\ and\ \bibinfo {author} {\bibfnamefont {T.~G.}\ \bibnamefont
  {Pedersen}},\ }\href {\doibase 10.1103/PhysRevB.80.195422} {\bibfield
  {journal} {\bibinfo  {journal} {Phys. Rev. B}\ }\textbf {\bibinfo {volume}
  {80}},\ \bibinfo {pages} {195422} (\bibinfo {year} {2009})}\BibitemShut
  {NoStop}%
\bibitem [{\citenamefont {Goupalov}\ \emph {et~al.}(2010)\citenamefont
  {Goupalov}, \citenamefont {Zarifi},\ and\ \citenamefont
  {Pedersen}}]{PhysRevB.81.153402}%
  \BibitemOpen
  \bibfield  {author} {\bibinfo {author} {\bibfnamefont {S.~V.}\ \bibnamefont
  {Goupalov}}, \bibinfo {author} {\bibfnamefont {A.}~\bibnamefont {Zarifi}}, \
  and\ \bibinfo {author} {\bibfnamefont {T.~G.}\ \bibnamefont {Pedersen}},\
  }\href {\doibase 10.1103/PhysRevB.81.153402} {\bibfield  {journal} {\bibinfo
  {journal} {Phys. Rev. B}\ }\textbf {\bibinfo {volume} {81}},\ \bibinfo
  {pages} {153402} (\bibinfo {year} {2010})}\BibitemShut {NoStop}%
\bibitem [{\citenamefont {Goupalov}(2005)}]{PhysRevB.72.195403}%
  \BibitemOpen
  \bibfield  {author} {\bibinfo {author} {\bibfnamefont {S.~V.}\ \bibnamefont
  {Goupalov}},\ }\href {\doibase 10.1103/PhysRevB.72.195403} {\bibfield
  {journal} {\bibinfo  {journal} {Phys. Rev. B}\ }\textbf {\bibinfo {volume}
  {72}},\ \bibinfo {pages} {195403} (\bibinfo {year} {2005})}\BibitemShut
  {NoStop}%
\bibitem [{\citenamefont {Berciaud}\ \emph {et~al.}(2010)\citenamefont
  {Berciaud}, \citenamefont {Voisin}, \citenamefont {Yan}, \citenamefont
  {Chandra}, \citenamefont {Caldwell}, \citenamefont {Shan}, \citenamefont
  {Brus}, \citenamefont {Hone},\ and\ \citenamefont
  {Heinz}}]{PhysRevB.81.041414}%
  \BibitemOpen
  \bibfield  {author} {\bibinfo {author} {\bibfnamefont {S.}~\bibnamefont
  {Berciaud}}, \bibinfo {author} {\bibfnamefont {C.}~\bibnamefont {Voisin}},
  \bibinfo {author} {\bibfnamefont {H.}~\bibnamefont {Yan}}, \bibinfo {author}
  {\bibfnamefont {B.}~\bibnamefont {Chandra}}, \bibinfo {author} {\bibfnamefont
  {R.}~\bibnamefont {Caldwell}}, \bibinfo {author} {\bibfnamefont
  {Y.}~\bibnamefont {Shan}}, \bibinfo {author} {\bibfnamefont {L.~E.}\
  \bibnamefont {Brus}}, \bibinfo {author} {\bibfnamefont {J.}~\bibnamefont
  {Hone}}, \ and\ \bibinfo {author} {\bibfnamefont {T.~F.}\ \bibnamefont
  {Heinz}},\ }\href {\doibase 10.1103/PhysRevB.81.041414} {\bibfield  {journal}
  {\bibinfo  {journal} {Phys. Rev. B}\ }\textbf {\bibinfo {volume} {81}},\
  \bibinfo {pages} {041414} (\bibinfo {year} {2010})}\BibitemShut {NoStop}%
\bibitem [{\citenamefont {Takagi}\ and\ \citenamefont
  {Okada}(2009)}]{PhysRevB.79.233406}%
  \BibitemOpen
  \bibfield  {author} {\bibinfo {author} {\bibfnamefont {Y.}~\bibnamefont
  {Takagi}}\ and\ \bibinfo {author} {\bibfnamefont {S.}~\bibnamefont {Okada}},\
  }\href {\doibase 10.1103/PhysRevB.79.233406} {\bibfield  {journal} {\bibinfo
  {journal} {Phys. Rev. B}\ }\textbf {\bibinfo {volume} {79}},\ \bibinfo
  {pages} {233406} (\bibinfo {year} {2009})}\BibitemShut {NoStop}%
\bibitem [{\citenamefont {Zarifi}\ and\ \citenamefont
  {Pedersen}(2006)}]{PhysRevB.74.155434}%
  \BibitemOpen
  \bibfield  {author} {\bibinfo {author} {\bibfnamefont {A.}~\bibnamefont
  {Zarifi}}\ and\ \bibinfo {author} {\bibfnamefont {T.~G.}\ \bibnamefont
  {Pedersen}},\ }\href {\doibase 10.1103/PhysRevB.74.155434} {\bibfield
  {journal} {\bibinfo  {journal} {Phys. Rev. B}\ }\textbf {\bibinfo {volume}
  {74}},\ \bibinfo {pages} {155434} (\bibinfo {year} {2006})}\BibitemShut
  {NoStop}%
\bibitem [{\citenamefont {Mali\ifmmode~\acute{c}\else \'{c}\fi{}}\ \emph
  {et~al.}(2006)\citenamefont {Mali\ifmmode~\acute{c}\else \'{c}\fi{}},
  \citenamefont {Hirtschulz}, \citenamefont {Milde}, \citenamefont {Knorr},\
  and\ \citenamefont {Reich}}]{PhysRevB.74.195431}%
  \BibitemOpen
  \bibfield  {author} {\bibinfo {author} {\bibfnamefont {E.}~\bibnamefont
  {Mali\ifmmode~\acute{c}\else \'{c}\fi{}}}, \bibinfo {author} {\bibfnamefont
  {M.}~\bibnamefont {Hirtschulz}}, \bibinfo {author} {\bibfnamefont
  {F.}~\bibnamefont {Milde}}, \bibinfo {author} {\bibfnamefont
  {A.}~\bibnamefont {Knorr}}, \ and\ \bibinfo {author} {\bibfnamefont
  {S.}~\bibnamefont {Reich}},\ }\href {\doibase 10.1103/PhysRevB.74.195431}
  {\bibfield  {journal} {\bibinfo  {journal} {Phys. Rev. B}\ }\textbf {\bibinfo
  {volume} {74}},\ \bibinfo {pages} {195431} (\bibinfo {year}
  {2006})}\BibitemShut {NoStop}%
\bibitem [{\citenamefont {Saito}\ \emph {et~al.}(2004)\citenamefont {Saito},
  \citenamefont {Gruneis}, \citenamefont {Samsonidze}, \citenamefont
  {Dresselhaus}, \citenamefont {Dresselhaus}, \citenamefont {Jorio},
  \citenamefont {Cancado}, \citenamefont {Pimenta},\ and\ \citenamefont
  {Souza}}]{ISI:000220192300002}%
  \BibitemOpen
  \bibfield  {author} {\bibinfo {author} {\bibfnamefont {R.}~\bibnamefont
  {Saito}}, \bibinfo {author} {\bibfnamefont {A.}~\bibnamefont {Gruneis}},
  \bibinfo {author} {\bibfnamefont {G.}~\bibnamefont {Samsonidze}}, \bibinfo
  {author} {\bibfnamefont {G.}~\bibnamefont {Dresselhaus}}, \bibinfo {author}
  {\bibfnamefont {M.}~\bibnamefont {Dresselhaus}}, \bibinfo {author}
  {\bibfnamefont {A.}~\bibnamefont {Jorio}}, \bibinfo {author} {\bibfnamefont
  {L.}~\bibnamefont {Cancado}}, \bibinfo {author} {\bibfnamefont
  {M.}~\bibnamefont {Pimenta}}, \ and\ \bibinfo {author} {\bibfnamefont
  {A.}~\bibnamefont {Souza}},\ }\href {\doibase 10.1007/s00339-003-2459-z}
  {\bibfield  {journal} {\bibinfo  {journal} {Applied Physics A}\ }\textbf
  {\bibinfo {volume} {78}},\ \bibinfo {pages} {1099} (\bibinfo {year}
  {2004})}\BibitemShut {NoStop}%
\bibitem [{\citenamefont {Miyauchi}\ \emph {et~al.}(2009)\citenamefont
  {Miyauchi}, \citenamefont {Hirori}, \citenamefont {Matsuda},\ and\
  \citenamefont {Kanemitsu}}]{PhysRevB.80.081410}%
  \BibitemOpen
  \bibfield  {author} {\bibinfo {author} {\bibfnamefont {Y.}~\bibnamefont
  {Miyauchi}}, \bibinfo {author} {\bibfnamefont {H.}~\bibnamefont {Hirori}},
  \bibinfo {author} {\bibfnamefont {K.}~\bibnamefont {Matsuda}}, \ and\
  \bibinfo {author} {\bibfnamefont {Y.}~\bibnamefont {Kanemitsu}},\ }\href
  {\doibase 10.1103/PhysRevB.80.081410} {\bibfield  {journal} {\bibinfo
  {journal} {Phys. Rev. B}\ }\textbf {\bibinfo {volume} {80}},\ \bibinfo
  {pages} {081410} (\bibinfo {year} {2009})}\BibitemShut {NoStop}%
\bibitem [{\citenamefont {Roche}\ \emph
  {et~al.}(2005{\natexlab{a}})\citenamefont {Roche}, \citenamefont {Jiang},
  \citenamefont {Triozon},\ and\ \citenamefont
  {Saito}}]{PhysRevLett.95.076803}%
  \BibitemOpen
  \bibfield  {author} {\bibinfo {author} {\bibfnamefont {S.}~\bibnamefont
  {Roche}}, \bibinfo {author} {\bibfnamefont {J.}~\bibnamefont {Jiang}},
  \bibinfo {author} {\bibfnamefont {F.~m.~c.}\ \bibnamefont {Triozon}}, \ and\
  \bibinfo {author} {\bibfnamefont {R.}~\bibnamefont {Saito}},\ }\href
  {\doibase 10.1103/PhysRevLett.95.076803} {\bibfield  {journal} {\bibinfo
  {journal} {Phys. Rev. Lett.}\ }\textbf {\bibinfo {volume} {95}},\ \bibinfo
  {pages} {076803} (\bibinfo {year} {2005}{\natexlab{a}})}\BibitemShut
  {NoStop}%
\bibitem [{\citenamefont {Tans}\ \emph {et~al.}(1997)\citenamefont {Tans},
  \citenamefont {Devoret}, \citenamefont {Dai}, \citenamefont {Thess},
  \citenamefont {Smalley}, \citenamefont {Geerligs},\ and\ \citenamefont
  {Dekker}}]{ISI:A1997WR25600045}%
  \BibitemOpen
  \bibfield  {author} {\bibinfo {author} {\bibfnamefont {S.}~\bibnamefont
  {Tans}}, \bibinfo {author} {\bibfnamefont {M.}~\bibnamefont {Devoret}},
  \bibinfo {author} {\bibfnamefont {H.}~\bibnamefont {Dai}}, \bibinfo {author}
  {\bibfnamefont {A.}~\bibnamefont {Thess}}, \bibinfo {author} {\bibfnamefont
  {R.}~\bibnamefont {Smalley}}, \bibinfo {author} {\bibfnamefont
  {L.}~\bibnamefont {Geerligs}}, \ and\ \bibinfo {author} {\bibfnamefont
  {C.}~\bibnamefont {Dekker}},\ }\href {\doibase 10.1038/386474a0} {\bibfield
  {journal} {\bibinfo  {journal} {Nature}\ }\textbf {\bibinfo {volume} {386}},\
  \bibinfo {pages} {474} (\bibinfo {year} {1997})}\BibitemShut {NoStop}%
\bibitem [{\citenamefont {Roche}\ \emph
  {et~al.}(2005{\natexlab{b}})\citenamefont {Roche}, \citenamefont {Jiang},
  \citenamefont {Triozon},\ and\ \citenamefont {Saito}}]{PhysRevB.72.113410}%
  \BibitemOpen
  \bibfield  {author} {\bibinfo {author} {\bibfnamefont {S.}~\bibnamefont
  {Roche}}, \bibinfo {author} {\bibfnamefont {J.}~\bibnamefont {Jiang}},
  \bibinfo {author} {\bibfnamefont {F.~m.~c.}\ \bibnamefont {Triozon}}, \ and\
  \bibinfo {author} {\bibfnamefont {R.}~\bibnamefont {Saito}},\ }\href
  {\doibase 10.1103/PhysRevB.72.113410} {\bibfield  {journal} {\bibinfo
  {journal} {Phys. Rev. B}\ }\textbf {\bibinfo {volume} {72}},\ \bibinfo
  {pages} {113410} (\bibinfo {year} {2005}{\natexlab{b}})}\BibitemShut
  {NoStop}%
\bibitem [{\citenamefont {Perebeinos}\ \emph {et~al.}(2005)\citenamefont
  {Perebeinos}, \citenamefont {Tersoff},\ and\ \citenamefont
  {Avouris}}]{PhysRevLett.94.086802}%
  \BibitemOpen
  \bibfield  {author} {\bibinfo {author} {\bibfnamefont {V.}~\bibnamefont
  {Perebeinos}}, \bibinfo {author} {\bibfnamefont {J.}~\bibnamefont {Tersoff}},
  \ and\ \bibinfo {author} {\bibfnamefont {P.}~\bibnamefont {Avouris}},\ }\href
  {\doibase 10.1103/PhysRevLett.94.086802} {\bibfield  {journal} {\bibinfo
  {journal} {Phys. Rev. Lett.}\ }\textbf {\bibinfo {volume} {94}},\ \bibinfo
  {pages} {086802} (\bibinfo {year} {2005})}\BibitemShut {NoStop}%
\bibitem [{\citenamefont {Bures}\ and\ \citenamefont {Ghosh}(1999)}]{Bures:99}%
  \BibitemOpen
  \bibfield  {author} {\bibinfo {author} {\bibfnamefont {J.}~\bibnamefont
  {Bures}}\ and\ \bibinfo {author} {\bibfnamefont {R.}~\bibnamefont {Ghosh}},\
  }\href {\doibase 10.1364/JOSAA.16.001992} {\bibfield  {journal} {\bibinfo
  {journal} {J. Opt. Soc. Am. A}\ }\textbf {\bibinfo {volume} {16}},\ \bibinfo
  {pages} {1992} (\bibinfo {year} {1999})}\BibitemShut {NoStop}%
\bibitem [{\citenamefont {Le~Kien}\ \emph {et~al.}(2005)\citenamefont
  {Le~Kien}, \citenamefont {Gupta}, \citenamefont {Nayak},\ and\ \citenamefont
  {Hakuta}}]{PhysRevA.72.063815}%
  \BibitemOpen
  \bibfield  {author} {\bibinfo {author} {\bibfnamefont {F.}~\bibnamefont
  {Le~Kien}}, \bibinfo {author} {\bibfnamefont {S.~D.}\ \bibnamefont {Gupta}},
  \bibinfo {author} {\bibfnamefont {K.~P.}\ \bibnamefont {Nayak}}, \ and\
  \bibinfo {author} {\bibfnamefont {K.}~\bibnamefont {Hakuta}},\ }\href
  {\doibase 10.1103/PhysRevA.72.063815} {\bibfield  {journal} {\bibinfo
  {journal} {Phys. Rev. A}\ }\textbf {\bibinfo {volume} {72}},\ \bibinfo
  {pages} {063815} (\bibinfo {year} {2005})}\BibitemShut {NoStop}%
\bibitem [{\citenamefont {Sagu\'e}\ \emph {et~al.}(2007)\citenamefont
  {Sagu\'e}, \citenamefont {Vetsch}, \citenamefont {Alt}, \citenamefont
  {Meschede},\ and\ \citenamefont {Rauschenbeutel}}]{PhysRevLett.99.163602}%
  \BibitemOpen
  \bibfield  {author} {\bibinfo {author} {\bibfnamefont {G.}~\bibnamefont
  {Sagu\'e}}, \bibinfo {author} {\bibfnamefont {E.}~\bibnamefont {Vetsch}},
  \bibinfo {author} {\bibfnamefont {W.}~\bibnamefont {Alt}}, \bibinfo {author}
  {\bibfnamefont {D.}~\bibnamefont {Meschede}}, \ and\ \bibinfo {author}
  {\bibfnamefont {A.}~\bibnamefont {Rauschenbeutel}},\ }\href {\doibase
  10.1103/PhysRevLett.99.163602} {\bibfield  {journal} {\bibinfo  {journal}
  {Phys. Rev. Lett.}\ }\textbf {\bibinfo {volume} {99}},\ \bibinfo {pages}
  {163602} (\bibinfo {year} {2007})}\BibitemShut {NoStop}%
\bibitem [{\citenamefont {Vetsch}\ \emph {et~al.}(2010)\citenamefont {Vetsch},
  \citenamefont {Reitz}, \citenamefont {Sagu\'e}, \citenamefont {Schmidt},
  \citenamefont {Dawkins},\ and\ \citenamefont
  {Rauschenbeutel}}]{PhysRevLett.104.203603}%
  \BibitemOpen
  \bibfield  {author} {\bibinfo {author} {\bibfnamefont {E.}~\bibnamefont
  {Vetsch}}, \bibinfo {author} {\bibfnamefont {D.}~\bibnamefont {Reitz}},
  \bibinfo {author} {\bibfnamefont {G.}~\bibnamefont {Sagu\'e}}, \bibinfo
  {author} {\bibfnamefont {R.}~\bibnamefont {Schmidt}}, \bibinfo {author}
  {\bibfnamefont {S.~T.}\ \bibnamefont {Dawkins}}, \ and\ \bibinfo {author}
  {\bibfnamefont {A.}~\bibnamefont {Rauschenbeutel}},\ }\href {\doibase
  10.1103/PhysRevLett.104.203603} {\bibfield  {journal} {\bibinfo  {journal}
  {Phys. Rev. Lett.}\ }\textbf {\bibinfo {volume} {104}},\ \bibinfo {pages}
  {203603} (\bibinfo {year} {2010})}\BibitemShut {NoStop}%
\bibitem [{\citenamefont {Liao}\ \emph {et~al.}(2008)\citenamefont {Liao},
  \citenamefont {Zhao},\ and\ \citenamefont {Pop}}]{PhysRevLett.101.256804}%
  \BibitemOpen
  \bibfield  {author} {\bibinfo {author} {\bibfnamefont {A.}~\bibnamefont
  {Liao}}, \bibinfo {author} {\bibfnamefont {Y.}~\bibnamefont {Zhao}}, \ and\
  \bibinfo {author} {\bibfnamefont {E.}~\bibnamefont {Pop}},\ }\href {\doibase
  10.1103/PhysRevLett.101.256804} {\bibfield  {journal} {\bibinfo  {journal}
  {Phys. Rev. Lett.}\ }\textbf {\bibinfo {volume} {101}},\ \bibinfo {pages}
  {256804} (\bibinfo {year} {2008})}\BibitemShut {NoStop}%
\bibitem [{\citenamefont {Dresselhaus}\ \emph {et~al.}(2007)\citenamefont
  {Dresselhaus}, \citenamefont {Dresselhaus}, \citenamefont {Saito},\ and\
  \citenamefont {Jorio}}]{annurev.physchem.58.032806.104628}%
  \BibitemOpen
  \bibfield  {author} {\bibinfo {author} {\bibfnamefont {M.~S.}\ \bibnamefont
  {Dresselhaus}}, \bibinfo {author} {\bibfnamefont {G.}~\bibnamefont
  {Dresselhaus}}, \bibinfo {author} {\bibfnamefont {R.}~\bibnamefont {Saito}},
  \ and\ \bibinfo {author} {\bibfnamefont {A.}~\bibnamefont {Jorio}},\ }\href
  {\doibase 10.1146/annurev.physchem.58.032806.104628} {\bibfield  {journal}
  {\bibinfo  {journal} {Annual Review of Physical Chemistry}\ }\textbf
  {\bibinfo {volume} {58}},\ \bibinfo {pages} {719} (\bibinfo {year}
  {2007})}\BibitemShut {NoStop}%
\bibitem [{\citenamefont {Bunder}\ and\ \citenamefont
  {Hill}(2009)}]{PhysRevB.80.153406}%
  \BibitemOpen
  \bibfield  {author} {\bibinfo {author} {\bibfnamefont {J.~E.}\ \bibnamefont
  {Bunder}}\ and\ \bibinfo {author} {\bibfnamefont {J.~M.}\ \bibnamefont
  {Hill}},\ }\href {\doibase 10.1103/PhysRevB.80.153406} {\bibfield  {journal}
  {\bibinfo  {journal} {Phys. Rev. B}\ }\textbf {\bibinfo {volume} {80}},\
  \bibinfo {pages} {153406} (\bibinfo {year} {2009})}\BibitemShut {NoStop}%
\bibitem [{\citenamefont {Mohite}\ \emph {et~al.}(2008)\citenamefont {Mohite},
  \citenamefont {Gopinath}, \citenamefont {Shah},\ and\ \citenamefont
  {Alphenaar}}]{doi:10.1021/nl0722525}%
  \BibitemOpen
  \bibfield  {author} {\bibinfo {author} {\bibfnamefont {A.~D.}\ \bibnamefont
  {Mohite}}, \bibinfo {author} {\bibfnamefont {P.}~\bibnamefont {Gopinath}},
  \bibinfo {author} {\bibfnamefont {H.~M.}\ \bibnamefont {Shah}}, \ and\
  \bibinfo {author} {\bibfnamefont {B.~W.}\ \bibnamefont {Alphenaar}},\ }\href
  {\doibase 10.1021/nl0722525} {\bibfield  {journal} {\bibinfo  {journal} {Nano
  Letters}\ }\textbf {\bibinfo {volume} {8}},\ \bibinfo {pages} {142} (\bibinfo
  {year} {2008})}\BibitemShut {NoStop}%
\bibitem [{\citenamefont {Ichida}\ \emph {et~al.}(1999)\citenamefont {Ichida},
  \citenamefont {Mizuno}, \citenamefont {Tani}, \citenamefont {Saito},\ and\
  \citenamefont {Nakamura}}]{JPSJ.68.3131}%
  \BibitemOpen
  \bibfield  {author} {\bibinfo {author} {\bibfnamefont {M.}~\bibnamefont
  {Ichida}}, \bibinfo {author} {\bibfnamefont {S.}~\bibnamefont {Mizuno}},
  \bibinfo {author} {\bibfnamefont {Y.}~\bibnamefont {Tani}}, \bibinfo {author}
  {\bibfnamefont {Y.}~\bibnamefont {Saito}}, \ and\ \bibinfo {author}
  {\bibfnamefont {A.}~\bibnamefont {Nakamura}},\ }\href {\doibase
  10.1143/JPSJ.68.3131} {\bibfield  {journal} {\bibinfo  {journal} {Journal of
  the Physical Society of Japan}\ }\textbf {\bibinfo {volume} {68}},\ \bibinfo
  {pages} {3131} (\bibinfo {year} {1999})}\BibitemShut {NoStop}%
\bibitem [{\citenamefont {Abergel}\ \emph {et~al.}(2010)\citenamefont
  {Abergel}, \citenamefont {Apalkov}, \citenamefont {Berashevich},
  \citenamefont {Ziegler},\ and\ \citenamefont
  {Chakraborty}}]{doi:10.1080/00018732.2010.487978}%
  \BibitemOpen
  \bibfield  {author} {\bibinfo {author} {\bibfnamefont {D.}~\bibnamefont
  {Abergel}}, \bibinfo {author} {\bibfnamefont {V.}~\bibnamefont {Apalkov}},
  \bibinfo {author} {\bibfnamefont {J.}~\bibnamefont {Berashevich}}, \bibinfo
  {author} {\bibfnamefont {K.}~\bibnamefont {Ziegler}}, \ and\ \bibinfo
  {author} {\bibfnamefont {T.}~\bibnamefont {Chakraborty}},\ }\href {\doibase
  10.1080/00018732.2010.487978} {\bibfield  {journal} {\bibinfo  {journal}
  {Advances in Physics}\ }\textbf {\bibinfo {volume} {59}},\ \bibinfo {pages}
  {261} (\bibinfo {year} {2010})}\BibitemShut {NoStop}%
\bibitem [{\citenamefont {Gr\"uneis}\ \emph {et~al.}(2003)\citenamefont
  {Gr\"uneis}, \citenamefont {Saito}, \citenamefont {Samsonidze}, \citenamefont
  {Kimura}, \citenamefont {Pimenta}, \citenamefont {Jorio}, \citenamefont
  {Filho}, \citenamefont {Dresselhaus},\ and\ \citenamefont
  {Dresselhaus}}]{PhysRevB.67.165402}%
  \BibitemOpen
  \bibfield  {author} {\bibinfo {author} {\bibfnamefont {A.}~\bibnamefont
  {Gr\"uneis}}, \bibinfo {author} {\bibfnamefont {R.}~\bibnamefont {Saito}},
  \bibinfo {author} {\bibfnamefont {G.~G.}\ \bibnamefont {Samsonidze}},
  \bibinfo {author} {\bibfnamefont {T.}~\bibnamefont {Kimura}}, \bibinfo
  {author} {\bibfnamefont {M.~A.}\ \bibnamefont {Pimenta}}, \bibinfo {author}
  {\bibfnamefont {A.}~\bibnamefont {Jorio}}, \bibinfo {author} {\bibfnamefont
  {A.~G.~S.}\ \bibnamefont {Filho}}, \bibinfo {author} {\bibfnamefont
  {G.}~\bibnamefont {Dresselhaus}}, \ and\ \bibinfo {author} {\bibfnamefont
  {M.~S.}\ \bibnamefont {Dresselhaus}},\ }\href {\doibase
  10.1103/PhysRevB.67.165402} {\bibfield  {journal} {\bibinfo  {journal} {Phys.
  Rev. B}\ }\textbf {\bibinfo {volume} {67}},\ \bibinfo {pages} {165402}
  (\bibinfo {year} {2003})}\BibitemShut {NoStop}%
\bibitem [{\citenamefont {Saito}\ \emph {et~al.}(2000)\citenamefont {Saito},
  \citenamefont {Dresselhaus},\ and\ \citenamefont
  {Dresselhaus}}]{PhysRevB.61.2981}%
  \BibitemOpen
  \bibfield  {author} {\bibinfo {author} {\bibfnamefont {R.}~\bibnamefont
  {Saito}}, \bibinfo {author} {\bibfnamefont {G.}~\bibnamefont {Dresselhaus}},
  \ and\ \bibinfo {author} {\bibfnamefont {M.~S.}\ \bibnamefont
  {Dresselhaus}},\ }\href {\doibase 10.1103/PhysRevB.61.2981} {\bibfield
  {journal} {\bibinfo  {journal} {Phys. Rev. B}\ }\textbf {\bibinfo {volume}
  {61}},\ \bibinfo {pages} {2981} (\bibinfo {year} {2000})}\BibitemShut
  {NoStop}%
\bibitem [{\citenamefont {Nikolaev}\ \emph {et~al.}(2009)\citenamefont
  {Nikolaev}, \citenamefont {Bibikov}, \citenamefont {Avdeenkov}, \citenamefont
  {Bodrenko},\ and\ \citenamefont {Tkalya}}]{PhysRevB.79.045418}%
  \BibitemOpen
  \bibfield  {author} {\bibinfo {author} {\bibfnamefont {A.~V.}\ \bibnamefont
  {Nikolaev}}, \bibinfo {author} {\bibfnamefont {A.~V.}\ \bibnamefont
  {Bibikov}}, \bibinfo {author} {\bibfnamefont {A.~V.}\ \bibnamefont
  {Avdeenkov}}, \bibinfo {author} {\bibfnamefont {I.~V.}\ \bibnamefont
  {Bodrenko}}, \ and\ \bibinfo {author} {\bibfnamefont {E.~V.}\ \bibnamefont
  {Tkalya}},\ }\href {\doibase 10.1103/PhysRevB.79.045418} {\bibfield
  {journal} {\bibinfo  {journal} {Phys. Rev. B}\ }\textbf {\bibinfo {volume}
  {79}},\ \bibinfo {pages} {045418} (\bibinfo {year} {2009})}\BibitemShut
  {NoStop}%
\bibitem [{\citenamefont {Blow}\ \emph {et~al.}(1990)\citenamefont {Blow},
  \citenamefont {Loudon}, \citenamefont {Phoenix},\ and\ \citenamefont
  {Shepherd}}]{PhysRevA.42.4102}%
  \BibitemOpen
  \bibfield  {author} {\bibinfo {author} {\bibfnamefont {K.~J.}\ \bibnamefont
  {Blow}}, \bibinfo {author} {\bibfnamefont {R.}~\bibnamefont {Loudon}},
  \bibinfo {author} {\bibfnamefont {S.~J.~D.}\ \bibnamefont {Phoenix}}, \ and\
  \bibinfo {author} {\bibfnamefont {T.~J.}\ \bibnamefont {Shepherd}},\ }\href
  {\doibase 10.1103/PhysRevA.42.4102} {\bibfield  {journal} {\bibinfo
  {journal} {Phys. Rev. A}\ }\textbf {\bibinfo {volume} {42}},\ \bibinfo
  {pages} {4102} (\bibinfo {year} {1990})}\BibitemShut {NoStop}%
\bibitem [{\citenamefont {Tasaki}\ \emph {et~al.}(1998)\citenamefont {Tasaki},
  \citenamefont {Maekawa},\ and\ \citenamefont {Yamabe}}]{PhysRevB.57.9301}%
  \BibitemOpen
  \bibfield  {author} {\bibinfo {author} {\bibfnamefont {S.}~\bibnamefont
  {Tasaki}}, \bibinfo {author} {\bibfnamefont {K.}~\bibnamefont {Maekawa}}, \
  and\ \bibinfo {author} {\bibfnamefont {T.}~\bibnamefont {Yamabe}},\ }\href
  {\doibase 10.1103/PhysRevB.57.9301} {\bibfield  {journal} {\bibinfo
  {journal} {Phys. Rev. B}\ }\textbf {\bibinfo {volume} {57}},\ \bibinfo
  {pages} {9301} (\bibinfo {year} {1998})}\BibitemShut {NoStop}%
\bibitem [{\citenamefont {Ajiki}\ and\ \citenamefont
  {Ando}(1994)}]{Ajiki1994349}%
  \BibitemOpen
  \bibfield  {author} {\bibinfo {author} {\bibfnamefont {H.}~\bibnamefont
  {Ajiki}}\ and\ \bibinfo {author} {\bibfnamefont {T.}~\bibnamefont {Ando}},\
  }\href {\doibase 10.1016/0921-4526(94)91112-6} {\bibfield  {journal}
  {\bibinfo  {journal} {Physica B: Condensed Matter}\ }\textbf {\bibinfo
  {volume} {201}},\ \bibinfo {pages} {349 } (\bibinfo {year}
  {1994})}\BibitemShut {NoStop}%
\bibitem [{\citenamefont {Islam}\ \emph {et~al.}(2004)\citenamefont {Islam},
  \citenamefont {Milkie}, \citenamefont {Kane}, \citenamefont {Yodh},\ and\
  \citenamefont {Kikkawa}}]{PhysRevLett.93.037404}%
  \BibitemOpen
  \bibfield  {author} {\bibinfo {author} {\bibfnamefont {M.~F.}\ \bibnamefont
  {Islam}}, \bibinfo {author} {\bibfnamefont {D.~E.}\ \bibnamefont {Milkie}},
  \bibinfo {author} {\bibfnamefont {C.~L.}\ \bibnamefont {Kane}}, \bibinfo
  {author} {\bibfnamefont {A.~G.}\ \bibnamefont {Yodh}}, \ and\ \bibinfo
  {author} {\bibfnamefont {J.~M.}\ \bibnamefont {Kikkawa}},\ }\href {\doibase
  10.1103/PhysRevLett.93.037404} {\bibfield  {journal} {\bibinfo  {journal}
  {Phys. Rev. Lett.}\ }\textbf {\bibinfo {volume} {93}},\ \bibinfo {pages}
  {037404} (\bibinfo {year} {2004})}\BibitemShut {NoStop}%
\bibitem [{\citenamefont {Lauret}\ \emph {et~al.}(2003)\citenamefont {Lauret},
  \citenamefont {Voisin}, \citenamefont {Cassabois}, \citenamefont {Delalande},
  \citenamefont {Roussignol}, \citenamefont {Jost},\ and\ \citenamefont
  {Capes}}]{PhysRevLett.90.057404}%
  \BibitemOpen
  \bibfield  {author} {\bibinfo {author} {\bibfnamefont {J.-S.}\ \bibnamefont
  {Lauret}}, \bibinfo {author} {\bibfnamefont {C.}~\bibnamefont {Voisin}},
  \bibinfo {author} {\bibfnamefont {G.}~\bibnamefont {Cassabois}}, \bibinfo
  {author} {\bibfnamefont {C.}~\bibnamefont {Delalande}}, \bibinfo {author}
  {\bibfnamefont {P.}~\bibnamefont {Roussignol}}, \bibinfo {author}
  {\bibfnamefont {O.}~\bibnamefont {Jost}}, \ and\ \bibinfo {author}
  {\bibfnamefont {L.}~\bibnamefont {Capes}},\ }\href {\doibase
  10.1103/PhysRevLett.90.057404} {\bibfield  {journal} {\bibinfo  {journal}
  {Phys. Rev. Lett.}\ }\textbf {\bibinfo {volume} {90}},\ \bibinfo {pages}
  {057404} (\bibinfo {year} {2003})}\BibitemShut {NoStop}%
\bibitem [{\citenamefont {Lin}(2000)}]{PhysRevB.62.13153}%
  \BibitemOpen
  \bibfield  {author} {\bibinfo {author} {\bibfnamefont {M.~F.}\ \bibnamefont
  {Lin}},\ }\href {\doibase 10.1103/PhysRevB.62.13153} {\bibfield  {journal}
  {\bibinfo  {journal} {Phys. Rev. B}\ }\textbf {\bibinfo {volume} {62}},\
  \bibinfo {pages} {13153} (\bibinfo {year} {2000})}\BibitemShut {NoStop}%
\bibitem [{\citenamefont {Tong}\ \emph {et~al.}(2003)\citenamefont {Tong},
  \citenamefont {Gattass}, \citenamefont {Ashcom}, \citenamefont {He},
  \citenamefont {Lou}, \citenamefont {Shen}, \citenamefont {Maxwell},\ and\
  \citenamefont {Mazur}}]{ISI:000187342000051}%
  \BibitemOpen
  \bibfield  {author} {\bibinfo {author} {\bibfnamefont {L.}~\bibnamefont
  {Tong}}, \bibinfo {author} {\bibfnamefont {R.}~\bibnamefont {Gattass}},
  \bibinfo {author} {\bibfnamefont {J.}~\bibnamefont {Ashcom}}, \bibinfo
  {author} {\bibfnamefont {S.}~\bibnamefont {He}}, \bibinfo {author}
  {\bibfnamefont {J.}~\bibnamefont {Lou}}, \bibinfo {author} {\bibfnamefont
  {M.}~\bibnamefont {Shen}}, \bibinfo {author} {\bibfnamefont {I.}~\bibnamefont
  {Maxwell}}, \ and\ \bibinfo {author} {\bibfnamefont {E.}~\bibnamefont
  {Mazur}},\ }\href {\doibase 10.1038/nature02193} {\bibfield  {journal}
  {\bibinfo  {journal} {Nature}\ }\textbf {\bibinfo {volume} {426}},\ \bibinfo
  {pages} {816} (\bibinfo {year} {2003})}\BibitemShut {NoStop}%
\bibitem [{\citenamefont {Tong}\ and\ \citenamefont
  {Mazur}(2008)}]{Tong20081240}%
  \BibitemOpen
  \bibfield  {author} {\bibinfo {author} {\bibfnamefont {L.}~\bibnamefont
  {Tong}}\ and\ \bibinfo {author} {\bibfnamefont {E.}~\bibnamefont {Mazur}},\
  }\href {\doibase 10.1016/j.jnoncrysol.2006.10.090} {\bibfield  {journal}
  {\bibinfo  {journal} {Journal of Non-Crystalline Solids}\ }\textbf {\bibinfo
  {volume} {354}},\ \bibinfo {pages} {1240 } (\bibinfo {year} {2008})},\
  \bibinfo {note} {proceedings of the 2005 International Conference on Glass -
  In conjunction with the Annual Meeting of the International Commission on
  Glass}\BibitemShut {NoStop}%
\bibitem [{\citenamefont {Brambilla}\ \emph {et~al.}(2004)\citenamefont
  {Brambilla}, \citenamefont {Finazzi},\ and\ \citenamefont
  {Richardson}}]{Brambilla:04}%
  \BibitemOpen
  \bibfield  {author} {\bibinfo {author} {\bibfnamefont {G.}~\bibnamefont
  {Brambilla}}, \bibinfo {author} {\bibfnamefont {V.}~\bibnamefont {Finazzi}},
  \ and\ \bibinfo {author} {\bibfnamefont {D.}~\bibnamefont {Richardson}},\
  }\href {\doibase 10.1364/OPEX.12.002258} {\bibfield  {journal} {\bibinfo
  {journal} {Opt. Express}\ }\textbf {\bibinfo {volume} {12}},\ \bibinfo
  {pages} {2258} (\bibinfo {year} {2004})}\BibitemShut {NoStop}%
\bibitem [{\citenamefont
  {Tong}(2010)}]{springerlink:10.1007/s12200-009-0073-1}%
  \BibitemOpen
  \bibfield  {author} {\bibinfo {author} {\bibfnamefont {L.}~\bibnamefont
  {Tong}},\ }\href {\doibase 10.1007/s12200-009-0073-1} {\bibfield  {journal}
  {\bibinfo  {journal} {Frontiers of Optoelectronics in China}\ }\textbf
  {\bibinfo {volume} {3}},\ \bibinfo {pages} {54} (\bibinfo {year}
  {2010})}\BibitemShut {NoStop}%
\bibitem [{\citenamefont {Tong}\ \emph {et~al.}(2004)\citenamefont {Tong},
  \citenamefont {Lou},\ and\ \citenamefont {Mazur}}]{ISI:000220594800008}%
  \BibitemOpen
  \bibfield  {author} {\bibinfo {author} {\bibfnamefont {L.}~\bibnamefont
  {Tong}}, \bibinfo {author} {\bibfnamefont {J.}~\bibnamefont {Lou}}, \ and\
  \bibinfo {author} {\bibfnamefont {E.}~\bibnamefont {Mazur}},\ }\href
  {\doibase 10.1364/OPEX.12.001025} {\bibfield  {journal} {\bibinfo  {journal}
  {Optics Express}\ }\textbf {\bibinfo {volume} {12}},\ \bibinfo {pages} {1025}
  (\bibinfo {year} {2004})}\BibitemShut {NoStop}%
\bibitem [{\citenamefont {Snyder}\ and\ \citenamefont
  {Love}(1983)}]{snyder1983optical}%
  \BibitemOpen
  \bibfield  {author} {\bibinfo {author} {\bibfnamefont {A.}~\bibnamefont
  {Snyder}}\ and\ \bibinfo {author} {\bibfnamefont {J.}~\bibnamefont {Love}},\
  }\href@noop {} {\emph {\bibinfo {title} {Optical waveguide theory}}},\
  Science paperbacks\ (\bibinfo  {publisher} {Chapman and Hall},\ \bibinfo
  {year} {1983})\BibitemShut {NoStop}%
\bibitem [{\citenamefont {Berciaud}\ \emph {et~al.}(2008)\citenamefont
  {Berciaud}, \citenamefont {Cognet},\ and\ \citenamefont
  {Lounis}}]{PhysRevLett.101.077402}%
  \BibitemOpen
  \bibfield  {author} {\bibinfo {author} {\bibfnamefont {S.}~\bibnamefont
  {Berciaud}}, \bibinfo {author} {\bibfnamefont {L.}~\bibnamefont {Cognet}}, \
  and\ \bibinfo {author} {\bibfnamefont {B.}~\bibnamefont {Lounis}},\ }\href
  {\doibase 10.1103/PhysRevLett.101.077402} {\bibfield  {journal} {\bibinfo
  {journal} {Phys. Rev. Lett.}\ }\textbf {\bibinfo {volume} {101}},\ \bibinfo
  {pages} {077402} (\bibinfo {year} {2008})}\BibitemShut {NoStop}%
\bibitem [{\citenamefont {Koskinen}(2010)}]{PhysRevB.82.193409}%
  \BibitemOpen
  \bibfield  {author} {\bibinfo {author} {\bibfnamefont {P.}~\bibnamefont
  {Koskinen}},\ }\href {\doibase 10.1103/PhysRevB.82.193409} {\bibfield
  {journal} {\bibinfo  {journal} {Phys. Rev. B}\ }\textbf {\bibinfo {volume}
  {82}},\ \bibinfo {pages} {193409} (\bibinfo {year} {2010})}\BibitemShut
  {NoStop}%
\bibitem [{\citenamefont {Ding}\ \emph {et~al.}(2009)\citenamefont {Ding},
  \citenamefont {Tselev}, \citenamefont {Wang}, \citenamefont {Yuan},
  \citenamefont {Chu}, \citenamefont {McNicholas}, \citenamefont {Li},\ and\
  \citenamefont {Liu}}]{doi:10.1021/nl803496s}%
  \BibitemOpen
  \bibfield  {author} {\bibinfo {author} {\bibfnamefont {L.}~\bibnamefont
  {Ding}}, \bibinfo {author} {\bibfnamefont {A.}~\bibnamefont {Tselev}},
  \bibinfo {author} {\bibfnamefont {J.}~\bibnamefont {Wang}}, \bibinfo {author}
  {\bibfnamefont {D.}~\bibnamefont {Yuan}}, \bibinfo {author} {\bibfnamefont
  {H.}~\bibnamefont {Chu}}, \bibinfo {author} {\bibfnamefont {T.~P.}\
  \bibnamefont {McNicholas}}, \bibinfo {author} {\bibfnamefont
  {Y.}~\bibnamefont {Li}}, \ and\ \bibinfo {author} {\bibfnamefont
  {J.}~\bibnamefont {Liu}},\ }\href {\doibase 10.1021/nl803496s} {\bibfield
  {journal} {\bibinfo  {journal} {Nano Letters}\ }\textbf {\bibinfo {volume}
  {9}},\ \bibinfo {pages} {800} (\bibinfo {year} {2009})}\BibitemShut {NoStop}%
\bibitem [{\citenamefont {Wang}\ \emph {et~al.}(2010)\citenamefont {Wang},
  \citenamefont {Ryu}, \citenamefont {De~Arco}, \citenamefont {Badmaev},
  \citenamefont {Zhang}, \citenamefont {Lin}, \citenamefont {Che},\ and\
  \citenamefont {Zhou}}]{springerlink:10.1007/s12274-010-0054-0}%
  \BibitemOpen
  \bibfield  {author} {\bibinfo {author} {\bibfnamefont {C.}~\bibnamefont
  {Wang}}, \bibinfo {author} {\bibfnamefont {K.}~\bibnamefont {Ryu}}, \bibinfo
  {author} {\bibfnamefont {L.}~\bibnamefont {De~Arco}}, \bibinfo {author}
  {\bibfnamefont {A.}~\bibnamefont {Badmaev}}, \bibinfo {author} {\bibfnamefont
  {J.}~\bibnamefont {Zhang}}, \bibinfo {author} {\bibfnamefont
  {X.}~\bibnamefont {Lin}}, \bibinfo {author} {\bibfnamefont {Y.}~\bibnamefont
  {Che}}, \ and\ \bibinfo {author} {\bibfnamefont {C.}~\bibnamefont {Zhou}},\
  }\href {\doibase 10.1007/s12274-010-0054-0} {\bibfield  {journal} {\bibinfo
  {journal} {Nano Research}\ }\textbf {\bibinfo {volume} {3}},\ \bibinfo
  {pages} {831} (\bibinfo {year} {2010})}\BibitemShut {NoStop}%
\bibitem [{\citenamefont {Sarker}\ \emph {et~al.}(2011)\citenamefont {Sarker},
  \citenamefont {Shekhar},\ and\ \citenamefont
  {Khondaker}}]{doi:10.1021/nn201314t}%
  \BibitemOpen
  \bibfield  {author} {\bibinfo {author} {\bibfnamefont {B.~K.}\ \bibnamefont
  {Sarker}}, \bibinfo {author} {\bibfnamefont {S.}~\bibnamefont {Shekhar}}, \
  and\ \bibinfo {author} {\bibfnamefont {S.~I.}\ \bibnamefont {Khondaker}},\
  }\href {\doibase 10.1021/nn201314t} {\bibfield  {journal} {\bibinfo
  {journal} {ACS Nano}\ }\textbf {\bibinfo {volume} {5}},\ \bibinfo {pages}
  {6297} (\bibinfo {year} {2011})}\BibitemShut {NoStop}%
\bibitem [{\citenamefont {Hong}\ \emph {et~al.}(2010)\citenamefont {Hong},
  \citenamefont {Banks},\ and\ \citenamefont {Rogers}}]{ADMA:ADMA200903238}%
  \BibitemOpen
  \bibfield  {author} {\bibinfo {author} {\bibfnamefont {S.~W.}\ \bibnamefont
  {Hong}}, \bibinfo {author} {\bibfnamefont {T.}~\bibnamefont {Banks}}, \ and\
  \bibinfo {author} {\bibfnamefont {J.~A.}\ \bibnamefont {Rogers}},\ }\href
  {\doibase 10.1002/adma.200903238} {\bibfield  {journal} {\bibinfo  {journal}
  {Advanced Materials}\ }\textbf {\bibinfo {volume} {22}},\ \bibinfo {pages}
  {1826} (\bibinfo {year} {2010})}\BibitemShut {NoStop}%
\bibitem [{\citenamefont {Shekhar}\ \emph {et~al.}(2011)\citenamefont
  {Shekhar}, \citenamefont {Stokes},\ and\ \citenamefont
  {Khondaker}}]{doi:10.1021/nn102305z}%
  \BibitemOpen
  \bibfield  {author} {\bibinfo {author} {\bibfnamefont {S.}~\bibnamefont
  {Shekhar}}, \bibinfo {author} {\bibfnamefont {P.}~\bibnamefont {Stokes}}, \
  and\ \bibinfo {author} {\bibfnamefont {S.~I.}\ \bibnamefont {Khondaker}},\
  }\href {\doibase 10.1021/nn102305z} {\bibfield  {journal} {\bibinfo
  {journal} {ACS Nano}\ }\textbf {\bibinfo {volume} {5}},\ \bibinfo {pages}
  {1739} (\bibinfo {year} {2011})}\BibitemShut {NoStop}%
\bibitem [{\citenamefont {Hadfield}(2009)}]{Hadfield}%
  \BibitemOpen
  \bibfield  {author} {\bibinfo {author} {\bibfnamefont {R.~H.}\ \bibnamefont
  {Hadfield}},\ }\href {\doibase 10.1038/nphoton.2009.23} {\bibfield  {journal}
  {\bibinfo  {journal} {Nature Photonics}\ }\textbf {\bibinfo {volume} {3}},\
  \bibinfo {pages} {696} (\bibinfo {year} {2009})}\BibitemShut {NoStop}%
\bibitem [{\citenamefont {Wang}\ \emph {et~al.}(2011)\citenamefont {Wang},
  \citenamefont {Luo}, \citenamefont {Sch\"{a}ffel}, \citenamefont
  {R\"{u}mmeli}, \citenamefont {Briggs},\ and\ \citenamefont
  {Warner}}]{0957-4484-22-24-245305}%
  \BibitemOpen
  \bibfield  {author} {\bibinfo {author} {\bibfnamefont {H.}~\bibnamefont
  {Wang}}, \bibinfo {author} {\bibfnamefont {J.}~\bibnamefont {Luo}}, \bibinfo
  {author} {\bibfnamefont {F.}~\bibnamefont {Sch\"{a}ffel}}, \bibinfo {author}
  {\bibfnamefont {M.~H.}\ \bibnamefont {R\"{u}mmeli}}, \bibinfo {author}
  {\bibfnamefont {G.~A.~D.}\ \bibnamefont {Briggs}}, \ and\ \bibinfo {author}
  {\bibfnamefont {J.~H.}\ \bibnamefont {Warner}},\ }\href {\doibase
  10.1088/0957-4484/22/24/245305} {\bibfield  {journal} {\bibinfo  {journal}
  {Nanotechnology}\ }\textbf {\bibinfo {volume} {22}},\ \bibinfo {pages}
  {245305} (\bibinfo {year} {2011})}\BibitemShut {NoStop}%
\bibitem [{\citenamefont {Xu}\ \emph {et~al.}(2006)\citenamefont {Xu},
  \citenamefont {Flor}, \citenamefont {Kim}, \citenamefont {Hamadani},
  \citenamefont {Schmidt}, \citenamefont {Smalley},\ and\ \citenamefont
  {Hauge}}]{doi:10.1021/ja060944+}%
  \BibitemOpen
  \bibfield  {author} {\bibinfo {author} {\bibfnamefont {Y.-Q.}\ \bibnamefont
  {Xu}}, \bibinfo {author} {\bibfnamefont {E.}~\bibnamefont {Flor}}, \bibinfo
  {author} {\bibfnamefont {M.~J.}\ \bibnamefont {Kim}}, \bibinfo {author}
  {\bibfnamefont {B.}~\bibnamefont {Hamadani}}, \bibinfo {author}
  {\bibfnamefont {H.}~\bibnamefont {Schmidt}}, \bibinfo {author} {\bibfnamefont
  {R.~E.}\ \bibnamefont {Smalley}}, \ and\ \bibinfo {author} {\bibfnamefont
  {R.~H.}\ \bibnamefont {Hauge}},\ }\href {\doibase 10.1021/ja060944+}
  {\bibfield  {journal} {\bibinfo  {journal} {Journal of the American Chemical
  Society}\ }\textbf {\bibinfo {volume} {128}},\ \bibinfo {pages} {6560}
  (\bibinfo {year} {2006})}\BibitemShut {NoStop}%
\bibitem [{\citenamefont {Hahm}\ \emph {et~al.}(2008)\citenamefont {Hahm},
  \citenamefont {Kwon}, \citenamefont {Lee}, \citenamefont {Ahn},\ and\
  \citenamefont {Jung}}]{doi:10.1021/jp8073877}%
  \BibitemOpen
  \bibfield  {author} {\bibinfo {author} {\bibfnamefont {M.~G.}\ \bibnamefont
  {Hahm}}, \bibinfo {author} {\bibfnamefont {Y.-K.}\ \bibnamefont {Kwon}},
  \bibinfo {author} {\bibfnamefont {E.}~\bibnamefont {Lee}}, \bibinfo {author}
  {\bibfnamefont {C.~W.}\ \bibnamefont {Ahn}}, \ and\ \bibinfo {author}
  {\bibfnamefont {Y.~J.}\ \bibnamefont {Jung}},\ }\href {\doibase
  10.1021/jp8073877} {\bibfield  {journal} {\bibinfo  {journal} {The Journal of
  Physical Chemistry C}\ }\textbf {\bibinfo {volume} {112}},\ \bibinfo {pages}
  {17143} (\bibinfo {year} {2008})}\BibitemShut {NoStop}%
\bibitem [{\citenamefont {Li}\ and\ \citenamefont {Zhang}(2011)}]{C1JM10399G}%
  \BibitemOpen
  \bibfield  {author} {\bibinfo {author} {\bibfnamefont {P.}~\bibnamefont
  {Li}}\ and\ \bibinfo {author} {\bibfnamefont {J.}~\bibnamefont {Zhang}},\
  }\href {\doibase 10.1039/C1JM10399G} {\bibfield  {journal} {\bibinfo
  {journal} {J. Mater. Chem.}\ }\textbf {\bibinfo {volume} {21}},\ \bibinfo
  {pages} {11815} (\bibinfo {year} {2011})}\BibitemShut {NoStop}%
\bibitem [{\citenamefont {Qu}\ \emph {et~al.}(2008)\citenamefont {Qu},
  \citenamefont {Du},\ and\ \citenamefont {Dai}}]{doi:10.1021/nl800967n}%
  \BibitemOpen
  \bibfield  {author} {\bibinfo {author} {\bibfnamefont {L.}~\bibnamefont
  {Qu}}, \bibinfo {author} {\bibfnamefont {F.}~\bibnamefont {Du}}, \ and\
  \bibinfo {author} {\bibfnamefont {L.}~\bibnamefont {Dai}},\ }\href {\doibase
  10.1021/nl800967n} {\bibfield  {journal} {\bibinfo  {journal} {Nano Letters}\
  }\textbf {\bibinfo {volume} {8}},\ \bibinfo {pages} {2682} (\bibinfo {year}
  {2008})}\BibitemShut {NoStop}%
\bibitem [{\citenamefont {Collins}\ \emph {et~al.}(2001)\citenamefont
  {Collins}, \citenamefont {Arnold},\ and\ \citenamefont
  {Avouris}}]{Collins27042001}%
  \BibitemOpen
  \bibfield  {author} {\bibinfo {author} {\bibfnamefont {P.~G.}\ \bibnamefont
  {Collins}}, \bibinfo {author} {\bibfnamefont {M.~S.}\ \bibnamefont {Arnold}},
  \ and\ \bibinfo {author} {\bibfnamefont {P.}~\bibnamefont {Avouris}},\ }\href
  {\doibase 10.1126/science.1058782} {\bibfield  {journal} {\bibinfo  {journal}
  {Science}\ }\textbf {\bibinfo {volume} {292}},\ \bibinfo {pages} {706}
  (\bibinfo {year} {2001})}\BibitemShut {NoStop}%
\bibitem [{\citenamefont {Zhong}\ \emph {et~al.}(2006)\citenamefont {Zhong},
  \citenamefont {Iwasaki},\ and\ \citenamefont {Kawarada}}]{Zhong20062009}%
  \BibitemOpen
  \bibfield  {author} {\bibinfo {author} {\bibfnamefont {G.~F.}\ \bibnamefont
  {Zhong}}, \bibinfo {author} {\bibfnamefont {T.}~\bibnamefont {Iwasaki}}, \
  and\ \bibinfo {author} {\bibfnamefont {H.}~\bibnamefont {Kawarada}},\ }\href
  {\doibase 10.1016/j.carbon.2006.01.027} {\bibfield  {journal} {\bibinfo
  {journal} {Carbon}\ }\textbf {\bibinfo {volume} {44}},\ \bibinfo {pages}
  {2009 } (\bibinfo {year} {2006})}\BibitemShut {NoStop}%
\bibitem [{\citenamefont {Murakami}\ \emph {et~al.}(2004)\citenamefont
  {Murakami}, \citenamefont {Chiashi}, \citenamefont {Miyauchi}, \citenamefont
  {Hu}, \citenamefont {Ogura}, \citenamefont {Okubo},\ and\ \citenamefont
  {Maruyama}}]{Murakami2004298}%
  \BibitemOpen
  \bibfield  {author} {\bibinfo {author} {\bibfnamefont {Y.}~\bibnamefont
  {Murakami}}, \bibinfo {author} {\bibfnamefont {S.}~\bibnamefont {Chiashi}},
  \bibinfo {author} {\bibfnamefont {Y.}~\bibnamefont {Miyauchi}}, \bibinfo
  {author} {\bibfnamefont {M.}~\bibnamefont {Hu}}, \bibinfo {author}
  {\bibfnamefont {M.}~\bibnamefont {Ogura}}, \bibinfo {author} {\bibfnamefont
  {T.}~\bibnamefont {Okubo}}, \ and\ \bibinfo {author} {\bibfnamefont
  {S.}~\bibnamefont {Maruyama}},\ }\href {\doibase
  10.1016/j.cplett.2003.12.095} {\bibfield  {journal} {\bibinfo  {journal}
  {Chemical Physics Letters}\ }\textbf {\bibinfo {volume} {385}},\ \bibinfo
  {pages} {298 } (\bibinfo {year} {2004})}\BibitemShut {NoStop}%
\bibitem [{\citenamefont {Esconjauregui}\ \emph {et~al.}(2010)\citenamefont
  {Esconjauregui}, \citenamefont {Fouquet}, \citenamefont {Bayer},
  \citenamefont {Ducati}, \citenamefont {Smajda}, \citenamefont {Hofmann},\
  and\ \citenamefont {Robertson}}]{doi:10.1021/nn1025675}%
  \BibitemOpen
  \bibfield  {author} {\bibinfo {author} {\bibfnamefont {S.}~\bibnamefont
  {Esconjauregui}}, \bibinfo {author} {\bibfnamefont {M.}~\bibnamefont
  {Fouquet}}, \bibinfo {author} {\bibfnamefont {B.~C.}\ \bibnamefont {Bayer}},
  \bibinfo {author} {\bibfnamefont {C.}~\bibnamefont {Ducati}}, \bibinfo
  {author} {\bibfnamefont {R.}~\bibnamefont {Smajda}}, \bibinfo {author}
  {\bibfnamefont {S.}~\bibnamefont {Hofmann}}, \ and\ \bibinfo {author}
  {\bibfnamefont {J.}~\bibnamefont {Robertson}},\ }\href {\doibase
  10.1021/nn1025675} {\bibfield  {journal} {\bibinfo  {journal} {ACS Nano}\
  }\textbf {\bibinfo {volume} {4}},\ \bibinfo {pages} {7431} (\bibinfo {year}
  {2010})}\BibitemShut {NoStop}%
\bibitem [{\citenamefont {Ibrahim}\ \emph {et~al.}(2011)\citenamefont
  {Ibrahim}, \citenamefont {Bachmatiuk}, \citenamefont {B\"{o}rrnert},
  \citenamefont {Bl\"{u}her}, \citenamefont {Wolff}, \citenamefont {Warner},
  \citenamefont {B\"{u}chner}, \citenamefont {Cuniberti},\ and\ \citenamefont
  {R\"{u}mmeli}}]{Ibrahim20115029}%
  \BibitemOpen
  \bibfield  {author} {\bibinfo {author} {\bibfnamefont {I.}~\bibnamefont
  {Ibrahim}}, \bibinfo {author} {\bibfnamefont {A.}~\bibnamefont {Bachmatiuk}},
  \bibinfo {author} {\bibfnamefont {F.}~\bibnamefont {B\"{o}rrnert}}, \bibinfo
  {author} {\bibfnamefont {J.}~\bibnamefont {Bl\"{u}her}}, \bibinfo {author}
  {\bibfnamefont {U.}~\bibnamefont {Wolff}}, \bibinfo {author} {\bibfnamefont
  {J.~H.}\ \bibnamefont {Warner}}, \bibinfo {author} {\bibfnamefont
  {B.}~\bibnamefont {B\"{u}chner}}, \bibinfo {author} {\bibfnamefont
  {G.}~\bibnamefont {Cuniberti}}, \ and\ \bibinfo {author} {\bibfnamefont
  {M.~H.}\ \bibnamefont {R\"{u}mmeli}},\ }\href {\doibase
  10.1016/j.carbon.2011.07.020} {\bibfield  {journal} {\bibinfo  {journal}
  {Carbon}\ }\textbf {\bibinfo {volume} {49}},\ \bibinfo {pages} {5029 }
  (\bibinfo {year} {2011})}\BibitemShut {NoStop}%
\bibitem [{\citenamefont {Wang}\ \emph {et~al.}(2004)\citenamefont {Wang},
  \citenamefont {Dukovic}, \citenamefont {Brus},\ and\ \citenamefont
  {Heinz}}]{PhysRevLett.92.177401}%
  \BibitemOpen
  \bibfield  {author} {\bibinfo {author} {\bibfnamefont {F.}~\bibnamefont
  {Wang}}, \bibinfo {author} {\bibfnamefont {G.}~\bibnamefont {Dukovic}},
  \bibinfo {author} {\bibfnamefont {L.~E.}\ \bibnamefont {Brus}}, \ and\
  \bibinfo {author} {\bibfnamefont {T.~F.}\ \bibnamefont {Heinz}},\ }\href
  {\doibase 10.1103/PhysRevLett.92.177401} {\bibfield  {journal} {\bibinfo
  {journal} {Phys. Rev. Lett.}\ }\textbf {\bibinfo {volume} {92}},\ \bibinfo
  {pages} {177401} (\bibinfo {year} {2004})}\BibitemShut {NoStop}%
\end{thebibliography}
\end{document}